# Combined Theory of Basis Sets of Spinors for Particles with Arbitrary Spin in Position, Momentum and Four-Dimensional Spaces

I.I.Guseinov

*Department of Physics, Faculty of Arts and Sciences, Onsekiz Mart University, Canakkale, Turkey*

**Abstract**

The 2(2s+1)-component relativistic basis spinors for the arbitrary spin particles are established in position, momentum and four-dimensional spaces, where $s = 0, 1/2, 1, 3/2, 2,...$. These spinors for integral- and half-integral spins are reduced to the independent sets of one- and two-component spinors, respectively. Relations presented in this study can be useful in the linear combination of atomic orbitals approximation for the solution of generalized Dirac equation of arbitrary spin particles introduced by the author when the orthogonal basis sets of relativistic exponential type spinor wave functions and Slater type spinor orbitals in position, momentum and four -dimensional spaces are employed.

## 1. Introduction

It is well known that the higher spin equations proposed by Dirac [1] have been developed and widely discussed in the literature [2-8]. All of these studies, which yield an adequate description of spin-s free particles, have many intrinsic contradictions and difficulties when an electromagnetic field interaction is introduced (see [9] and references therein). The generalized Dirac equations presented in our previous works [10] and [11] for the particles with half-integral and arbitrary spins, respectively, are consistent and causal in the presence of an electromagnetic field interaction.

The elaboration of algorithms for the solution of the generalized Dirac equation for the particles with arbitrary spin [11] in linear combination of atomic orbitals (LCAO) approach [12-14] necessitates progress in the development of theory for complete orthonormal basis sets of relativistic spinors of multiple orders. The method for constructing in position, momentum and four-dimensional spaces the complete orthonormal basis sets for (2s+1)-component relativistic tensor wave functions and Slater tensor orbitals has been suggested in previous article [15]. Extending this approach to the case of spinors of multiple order and using the method set out in [16], we construct in this study the relevant complete orthonormal basis sets of 2(2s+1)-component relativistic $\Psi^{\alpha s}$-exponential type spinor orbitals ($\Psi^{\alpha s}$-ETSO) for particles with arbitrary spin in position, momentum and four-dimensional spaces through the sets of one- and two-component spinor type tensor spherical harmonics and radial parts of the complete orthonormal sets of nonrelativistic $\psi^\alpha$-exponential type orbitals ($\psi^\alpha$-ETO) [17] the angular parts of which are the scalar spherical harmonics. The indices $\alpha$ occurring in the radial parts of $\psi^\alpha$-ETO is the frictional quantum number [18]. It should be noted that the nonrelativistic $\psi^\alpha$-ETO are the special cases of $\Psi^{\alpha s}$-ETSO for s=0, i.e., $\Psi^{\alpha 0} \equiv \psi^\alpha$.

The basis sets of relativistic spinors of multiple order obtained might be useful for solution of generalized Dirac equation of arbitrary spin particles when the complete orthonormal relativistic $\Psi^{\alpha s}$-ETSO basis sets in LCAO approximation are employed. We notice that the definition of phases in this work for the scalar spherical harmonics $\left(Y^*_{lm_l} = Y_{l-m_l}\right)$ differs from the Condon-Shortley phases [19] by the sing factor $\left(Y^*_{lm_l} = i^{|m_l|+m_l} Y_{l-m_l}\right)$.

## 2. Relativistic spinor type tensor spherical harmonics

In order to construct the complete orthonormal basis sets of relativistic $\Psi^{\alpha s}$-ETSO and $X^s$-Slater type spinor orbitals ($X^s$-STSO) of 2(2s+1) order in position, momentum and four-dimensional spaces we introduce the following formulae for the independent spinor type tensor (STT) spherical harmonics of (2s+1) order (see Ref. [15]):

for integral spin:

$$H^s_{ljm}(\theta,\varphi) = \begin{bmatrix} H^{s0}_{ljm}(\theta,\varphi) \\ H^{s1}_{ljm}(\theta,\varphi) \\ \vdots \\ H^{s,2s-1}_{ljm}(\theta,\varphi) \\ H^{s,2s}_{ljm}(\theta,\varphi) \end{bmatrix} \quad (1a)$$

$$\mathcal{H}^s_{ljm}(\theta,\varphi) = \begin{bmatrix} \mathcal{H}^{s,2s}_{ljm}(\theta,\varphi) \\ \mathcal{H}^{s,2s-1}_{ljm}(\theta,\varphi) \\ \vdots \\ \mathcal{H}^{s1}_{ljm}(\theta,\varphi) \\ \mathcal{H}^{s0}_{ljm}(\theta,\varphi) \end{bmatrix} \quad (1b)$$

for half-integral spin

$$Y_{ljm}^{s}(\theta,\varphi) = \begin{bmatrix} Y_{ljm}^{s0}(\theta,\varphi) \\ Y_{ljm}^{s2}(\theta,\varphi) \\ \vdots \\ Y_{ljm}^{s,2s-3}(\theta,\varphi) \\ Y_{ljm}^{s,2s-1}(\theta,\varphi) \end{bmatrix} \quad (2a)$$

$$\Upsilon_{ljm}^{s}(\theta,\varphi) = \begin{bmatrix} \Upsilon_{ljm}^{s,2s-1}(\theta,\varphi) \\ \Upsilon_{ljm}^{s,2s-3}(\theta,\varphi) \\ \vdots \\ \Upsilon_{ljm}^{s2}(\theta,\varphi) \\ \Upsilon_{ljm}^{s0}(\theta,\varphi) \end{bmatrix} \quad (2b)$$

These STT spherical harmonics are eigenfunctions of operators $\hat{j}^2, \hat{j}_z, \hat{l}^2$ and $\hat{s}^2$. The one- and two-component basis sets of STT spherical harmonics $H_{ljm}^{s\lambda}(\theta,\varphi)$, $\mathcal{H}_{ljm}^{s\lambda}(\theta,\varphi)$ and $Y_{ljm}^{s\lambda}(\theta,\varphi)$, $\Upsilon_{ljm}^{s\lambda}(\theta,\varphi)$ occurring in Eqs. (1a), (1b) and (2a), (2b), respectively, can be expressed through the scalar spherical harmonics:

$$H_{ljm}^{s\lambda}(\theta,\varphi) = a_{ljm}^{s}(\lambda)\beta_{m(\lambda)}Y_{lm(\lambda)}(\theta,\varphi) \quad (3a)$$

$$\mathcal{H}_{ljm}^{s\lambda}(\theta,\varphi) = -ia_{ljm}^{s}(\lambda)\beta_{m(\lambda)}Y_{lm(\lambda)}(\theta,\varphi) \quad (3b)$$

$$Y_{ljm}^{s\lambda}(\theta,\varphi) = \begin{bmatrix} a_{ljm}^{s}(\lambda)\beta_{m(\lambda)}Y_{lm(\lambda)}(\theta,\varphi) \\ a_{ljm}^{s}(\lambda+1)\beta_{m(\lambda+1)}Y_{lm(\lambda+1)}(\theta,\varphi) \end{bmatrix} \quad (4a)$$

$$\Upsilon_{ljm}^{s\lambda}(\theta,\varphi) = \begin{bmatrix} -ia_{ljm}^{s}(\lambda+1)\beta_{m(\lambda+1)}Y_{lm(\lambda+1)}(\theta,\varphi) \\ -ia_{ljm}^{s}(\lambda)\beta_{m(\lambda)}Y_{lm(\lambda)}(\theta,\varphi) \end{bmatrix}, \quad (4b)$$

where

for integral spin

$$0 \leq \lambda(1) \leq 2s, \quad |l-s| \leq j \leq j+s, \; -j \leq m \leq j, \quad j = l + \frac{1}{2}t, \quad t = 2(j-l) = 0, \pm 2, \ldots, \pm 2s,$$

$m_l = m(\lambda) = m - s + \lambda$ and $\beta_{m(\lambda)} = (-1)^{[|m(\lambda)|-m(\lambda)]/2}$,

for half-integral spin

$$0 \leq \lambda(2) \leq 2s-1, \; |l-s| \leq j \leq j+s, \; -j \leq m \leq j, \; j = l + \frac{1}{2}t, \quad t = 2(j-l) = \pm 1, \pm 3, \ldots, \pm 2s,$$

$m_l = m(\lambda) = m - s + \lambda, \quad \beta_{m(\lambda)} = (-1)^{[|m(\lambda)|-m(\lambda)]/2}$.

Here, $a_{ljm}^{s}(\lambda)$ are the modified Clebsch-Gordan coefficients defined as

$$a_{ljm}^{s}(\lambda) = \left(lsm(\lambda)s - \lambda | lsjm\right). \tag{5}$$

See Ref. [19] for the definition of Clebsch-Gordan coefficients $\left(lsm_l m - m_l | lsjm\right)$.

The STT spherical harmonics $H_{ljm}^{s}(\theta,\varphi)$, $\mathcal{H}_{ljm}^{s}(\theta,\varphi)$ and $Y_{ljm}^{s}(\theta,\varphi)$, $\Upsilon_{ljm}^{s}(\theta,\varphi)$ for fixed s satisfy the following orthonormality relations:

$$\int_{0}^{\pi}\int_{0}^{2\pi} H_{ljm}^{s*}(\theta,\varphi) H_{l'j'm'}^{s}(\theta,\varphi) \sin\theta d\theta d\varphi = \sum_{\lambda=0}^{2s}\int_{0}^{\pi}\int_{0}^{2\pi} H_{ljm}^{s\lambda*}(\theta,\varphi) H_{l'j'm'}^{s\lambda}(\theta,\varphi) \sin\theta d\theta d\varphi = \delta_{ll'}\delta_{jj'}\delta_{mm'} \tag{6a}$$

$$\int_{0}^{\pi}\int_{0}^{2\pi} \mathcal{H}_{ljm}^{s*}(\theta,\varphi) \mathcal{H}_{l'j'm'}^{s}(\theta,\varphi) \sin\theta d\theta d\varphi = \sum_{\lambda=0}^{2s}\int_{0}^{\pi}\int_{0}^{2\pi} \mathcal{H}_{ljm}^{s\lambda*}(\theta,\varphi) \mathcal{H}_{l'j'm'}^{s\lambda}(\theta,\varphi) \sin\theta d\theta d\varphi = \delta_{ll'}\delta_{jj'}\delta_{mm'} \tag{6b}$$

$$\int_{0}^{\pi}\int_{0}^{2\pi} Y_{ljm}^{s\dagger}(\theta,\varphi) Y_{l'j'm'}^{s}(\theta,\varphi) \sin\theta d\theta d\varphi = \sum_{\lambda=0}^{2s-1}\int_{0}^{\pi}\int_{0}^{2\pi} Y_{ljm}^{s\lambda\dagger}(\theta,\varphi) Y_{l'j'm'}^{s\lambda}(\theta,\varphi) \sin\theta d\theta d\varphi = \delta_{ll'}\delta_{jj'}\delta_{mm'} \tag{7a}$$

$$\int_{0}^{\pi}\int_{0}^{2\pi} \Upsilon_{ljm}^{s\dagger}(\theta,\varphi) \Upsilon_{l'j'm'}^{s}(\theta,\varphi) \sin\theta d\theta d\varphi = \sum_{\lambda=0}^{2s-1}\int_{0}^{\pi}\int_{0}^{2\pi} \Upsilon_{ljm}^{s\lambda\dagger}(\theta,\varphi) \Upsilon_{l'j'm'}^{s\lambda}(\theta,\varphi) \sin\theta d\theta d\varphi = \delta_{ll'}\delta_{jj'}\delta_{mm'}. \tag{7b}$$

## 3. Basis sets of relativistic $\Psi^{\alpha s}$-ETSO and $X^s$-STSO functions

To construct the basis sets of 2(2s+1)-component relativistic spinors from STT spherical harmonics and radial parts of nonrelativistic orbitals we use the method set out in a previous paper [10]. Then, we obtain for the complete basis sets of relativistic spinor wave functions $\Psi^{\alpha s}$, $\bar{\Psi}^{\alpha s}$ and Slater spinor orbitals $X^s$ in position space the following relations:

for integral spin

$$\Psi_{nljm}^{\alpha s}(r,\theta,\varphi) = \frac{1}{\sqrt{2}}\begin{bmatrix} R_{nl}^{\alpha}(r) H_{ljm}^{s}(\theta,\varphi) \\ \tilde{R}_{\tilde{n}l}^{\alpha}(r) \mathcal{H}_{ljm}^{s}(\theta,\varphi) \end{bmatrix} = \frac{1}{\sqrt{2}}\begin{bmatrix} a_{ljm}^{s}(0)\beta_{m(0)}\psi_{nlm(0)}^{\alpha}(r,\theta,\varphi) \\ a_{ljm}^{s}(1)\beta_{m(1)}\psi_{nlm(1)}^{\alpha}(r,\theta,\varphi) \\ \vdots \\ a_{ljm}^{s}(2s-1)\beta_{m(2s-1)}\psi_{nlm(2s-1)}^{\alpha}(r,\theta,\varphi) \\ a_{ljm}^{s}(2s)\beta_{m(2s)}\psi_{nlm(2s)}^{\alpha}(r,\theta,\varphi) \\ -ia_{ljm}^{s}(2s)\beta_{m(2s)}\psi_{\tilde{n}lm(2s)}^{\alpha}(r,\theta,\varphi) \\ -ia_{ljm}^{s}(2s-1)\beta_{m(2s-1)}\psi_{\tilde{n}lm(2s-1)}^{\alpha}(r,\theta,\varphi) \\ \vdots \\ -ia_{ljm}^{s}(1)\beta_{m(1)}\psi_{\tilde{n}lm(1)}^{\alpha}(r,\theta,\varphi) \\ -ia_{ljm}^{s}(0)\beta_{m(0)}\psi_{\tilde{n}lm(0)}^{\alpha}(r,\theta,\varphi) \end{bmatrix} \tag{8a}$$

$$\overline{\Psi}_{nljm}^{\alpha s}(r,\theta,\varphi) = \frac{1}{\sqrt{2}}\begin{bmatrix} \overline{R}_{nl}^{\alpha}(r) H_{ljm}^{s}(\theta,\varphi) \\ \tilde{\overline{R}}_{\tilde{n}l}^{\alpha}(r) \mathcal{H}_{ljm}^{s}(\theta,\varphi) \end{bmatrix} = \frac{1}{\sqrt{2}}\begin{bmatrix} a_{ljm}^{s}(0)\beta_{m(0)}\overline{\psi}_{nlm(0)}^{\alpha}(r,\theta,\varphi) \\ a_{ljm}^{s}(1)\beta_{m(1)}\overline{\psi}_{nlm(1)}^{\alpha}(r,\theta,\varphi) \\ \vdots \\ a_{ljm}^{s}(2s-1)\beta_{m(2s-1)}\overline{\psi}_{nlm(2s-1)}^{\alpha}(r,\theta,\varphi) \\ a_{ljm}^{s}(2s)\beta_{m(2s)}\overline{\psi}_{nlm(2s)}^{\alpha}(r,\theta,\varphi) \\ -ia_{ljm}^{s}(2s)\beta_{m(2s)}\overline{\psi}_{\tilde{n}lm(2s)}^{\alpha}(r,\theta,\varphi) \\ -ia_{ljm}^{s}(2s-1)\beta_{m(2s-1)}\overline{\psi}_{\tilde{n}lm(2s-1)}^{\alpha}(r,\theta,\varphi) \\ \vdots \\ -ia_{ljm}^{s}(1)\beta_{m(1)}\overline{\psi}_{\tilde{n}lm(1)}^{\alpha}(r,\theta,\varphi) \\ -ia_{ljm}^{s}(0)\beta_{m(0)}\overline{\psi}_{\tilde{n}lm(0)}^{\alpha}(r,\theta,\varphi) \end{bmatrix} \quad (8b)$$

$$X_{nljm}^{s}(r,\theta,\varphi) = \frac{1}{\sqrt{2}}\begin{bmatrix} R_{n}(r) H_{ljm}^{s}(\theta,\varphi) \\ R_{n}(r) \mathcal{H}_{ljm}^{s}(\theta,\varphi) \end{bmatrix} = \frac{1}{\sqrt{2}}\begin{bmatrix} a_{ljm}^{s}(0)\beta_{m(0)}\chi_{nlm(0)}(r,\theta,\varphi) \\ a_{ljm}^{s}(1)\beta_{m(1)}\chi_{nlm(1)}(r,\theta,\varphi) \\ \vdots \\ a_{ljm}^{s}(2s-1)\beta_{m(2s-1)}\chi_{nlm(2s-1)}(r,\theta,\varphi) \\ a_{ljm}^{s}(2s)\beta_{m(2s)}\chi_{nlm(2s)}(r,\theta,\varphi) \\ -ia_{ljm}^{s}(2s)\beta_{m(2s)}\chi_{\tilde{n}lm(2s)}(r,\theta,\varphi) \\ -ia_{ljm}^{s}(2s-1)\beta_{m(2s-1)}\chi_{\tilde{n}lm(2s-1)}(r,\theta,\varphi) \\ \vdots \\ -ia_{ljm}^{s}(1)\beta_{m(1)}\chi_{\tilde{n}lm(1)}(r,\theta,\varphi) \\ -ia_{ljm}^{s}(0)\beta_{m(0)}\chi_{\tilde{n}lm(0)}(r,\theta,\varphi) \end{bmatrix}, \quad (9)$$

for half-integral spin

$$\Psi_{nljm}^{\alpha s}(r,\theta,\varphi) = \frac{1}{\sqrt{2}}\begin{bmatrix} R_{nl}^{\alpha}(r) Y_{ljm}^{s}(\theta,\varphi) \\ R_{\tilde{n}l}^{\alpha}(r) \Upsilon_{ljm}^{s}(\theta,\varphi) \end{bmatrix} = \frac{1}{\sqrt{2}}\begin{bmatrix} a_{ljm}^{s}(0)\beta_{m(0)}\psi_{nlm(0)}^{\alpha}(r,\theta,\varphi) \\ a_{ljm}^{s}(2)\beta_{m(2)}\psi_{nlm(2)}^{\alpha}(r,\theta,\varphi) \\ \vdots \\ a_{ljm}^{s}(2s-3)\beta_{m(2s-3)}\psi_{nlm(2s-3)}^{\alpha}(r,\theta,\varphi) \\ a_{ljm}^{s}(2s-1)\beta_{m(2s-1)}\psi_{nlm(2s-1)}^{\alpha}(r,\theta,\varphi) \\ -ia_{ljm}^{s}(2s-1)\beta_{m(2s-1)}\psi_{\tilde{n}lm(2s-1)}^{\alpha}(r,\theta,\varphi) \\ -ia_{ljm}^{s}(2s-3)\beta_{m(2s-3)}\psi_{\tilde{n}lm(2s-3)}^{\alpha}(r,\theta,\varphi) \\ \vdots \\ -ia_{ljm}^{s}(2)\beta_{m(2)}\psi_{\tilde{n}lm(2)}^{\alpha}(r,\theta,\varphi) \\ -ia_{ljm}^{s}(0)\beta_{m(0)}\psi_{\tilde{n}lm(0)}^{\alpha}(r,\theta,\varphi) \end{bmatrix} \quad (10a)$$

$$\overline{\Psi}_{nljm}^{\alpha s}(r,\theta,\varphi) = \frac{1}{\sqrt{2}}\begin{bmatrix} \overline{R}_{nl}^{\alpha}(r)Y_{ljm}^{s}(\theta,\varphi) \\ \tilde{\overline{R}}_{\tilde{n}l}^{\alpha}(r)\Upsilon_{ljm}^{s}(\theta,\varphi) \end{bmatrix} = \frac{1}{\sqrt{2}}\begin{bmatrix} a_{ljm}^{s}(0)\beta_{m(0)}\overline{\psi}_{nlm(0)}^{\alpha}(r,\theta,\varphi) \\ a_{ljm}^{s}(2)\beta_{m(2)}\overline{\psi}_{nlm(2)}^{\alpha}(r,\theta,\varphi) \\ \vdots \\ a_{ljm}^{s}(2s-3)\beta_{m(2s-3)}\overline{\psi}_{nlm(2s-3)}^{\alpha}(r,\theta,\varphi) \\ a_{ljm}^{s}(2s-1)\beta_{m(2s-1)}\overline{\psi}_{nlm(2s-1)}^{\alpha}(r,\theta,\varphi) \\ -ia_{ljm}^{s}(2s-1)\beta_{m(2s-1)}\overline{\psi}_{\tilde{n}lm(2s-1)}^{\alpha}(r,\theta,\varphi) \\ -ia_{ljm}^{s}(2s-3)\beta_{m(2s-3)}\overline{\psi}_{\tilde{n}lm(2s-3)}^{\alpha}(r,\theta,\varphi) \\ \vdots \\ -ia_{ljm}^{s}(2)\beta_{m(2)}\overline{\psi}_{\tilde{n}lm(2)}^{\alpha}(r,\theta,\varphi) \\ -ia_{ljm}^{s}(0)\beta_{m(0)}\overline{\psi}_{\tilde{n}lm(0)}^{\alpha}(r,\theta,\varphi) \end{bmatrix} \quad (10b)$$

$$X_{nljm}^{s}(r,\theta,\varphi) = \frac{1}{\sqrt{2}}\begin{bmatrix} R_{n}(r)Y_{ljm}^{s}(\theta,\varphi) \\ \tilde{R}_{\tilde{n}}(r)\Upsilon_{ljm}^{s}(\theta,\varphi) \end{bmatrix} = \frac{1}{\sqrt{2}}\begin{bmatrix} a_{ljm}^{s}(0)\beta_{m(0)}\chi_{nlm(0)}(r,\theta,\varphi) \\ a_{ljm}^{s}(2)\beta_{m(2)}\chi_{nlm(2)}(r,\theta,\varphi) \\ \vdots \\ a_{ljm}^{s}(2s-3)\beta_{m(2s-3)}\chi_{nlm(2s-3)}(r,\theta,\varphi) \\ a_{ljm}^{s}(2s-1)\beta_{m(2s-1)}\chi_{nlm(2s-1)}(r,\theta,\varphi) \\ -ia_{ljm}^{s}(2s-1)\beta_{m(2s-1)}\chi_{\tilde{n}lm(2s-1)}(r,\theta,\varphi) \\ -ia_{ljm}^{s}(2s-3)\beta_{m(2s-3)}\chi_{\tilde{n}lm(2s-3)}(r,\theta,\varphi) \\ \vdots \\ -ia_{ljm}^{s}(2)\beta_{m(2)}\chi_{\tilde{n}lm(2)}(r,\theta,\varphi) \\ -ia_{ljm}^{s}(0)\beta_{m(0)}\chi_{\tilde{n}lm(0)}(r,\theta,\varphi) \end{bmatrix}, \quad (11)$$

where $n \geq 1$, $s \leq j \leq s+n-1$, $j-s \leq l \leq \min(j+s, n-1)$ and

$$\tilde{R}_{\tilde{n}}(r) = R_{\tilde{n}}(\zeta,r) = \frac{(2\zeta)^{\tilde{n}+\frac{1}{2}}}{\sqrt{(2\tilde{n})!}}r^{\tilde{n}-1}e^{-\zeta r}. \quad (12)$$

The relativistic spinor wave functions ($K_{nljm}^{\alpha s}, \overline{K}_{nljm}^{\alpha s}$) and Slater spinor orbitals $K_{nljm}^{s}$ in position, momentum and four-dimensional spaces are defined as

$$K_{nljm}^{\alpha s} \equiv \Psi_{nljm}^{\alpha s}(\zeta,\vec{r}), \Phi_{nljm}^{\alpha s}(\zeta,\vec{k}), Z_{nljm}^{\alpha s}(\zeta,\beta\theta\varphi) \quad (13)$$

$$\overline{K}_{nljm}^{\alpha s} \equiv \overline{\Psi}_{nljm}^{\alpha s}(\zeta,\vec{r}), \overline{\Phi}_{nljm}^{\alpha s}(\zeta,\vec{k}), \overline{Z}_{nljm}^{\alpha s}(\zeta,\beta\theta\varphi) \quad (14)$$

$$K_{nljm}^{s} \equiv X_{nljm}^{s}(\zeta,\vec{r}), U_{nljm}^{s}(\zeta,\vec{k}), V_{nljm}^{s}(\zeta,\beta\theta\varphi). \quad (15)$$

Here, the $k_{nlm(\lambda)}^{\alpha}$, $\overline{k}_{nlm(\lambda)}^{\alpha}$ and $k_{nlm(\lambda)}$ are the nonrelativistic complete basis sets of orbitals. They are determined through the corresponding nonrelativistic functions in position, momentum and four-dimensional spaces by

$$k^{\alpha}_{nlm(\lambda)} \equiv \psi^{\alpha}_{nlm(\lambda)}(\zeta,\vec{r}), \phi^{\alpha}_{nlm(\lambda)}(\zeta,\vec{k}), z^{\alpha}_{nlm(\lambda)}(\zeta,\beta\theta\varphi) \tag{16}$$

$$\overline{k}^{\alpha}_{nlm(\lambda)} \equiv \overline{\psi}^{\alpha}_{nlm(\lambda)}(\zeta,\vec{r}), \overline{\phi}^{\alpha}_{nlm(\lambda)}(\zeta,\vec{k}), \overline{z}^{\alpha}_{nlm(\lambda)}(\zeta,\beta\theta\varphi) \tag{17}$$

$$k_{nlm(\lambda)} \equiv \chi_{nlm(\lambda)}(\zeta,\vec{r}), u_{nlm(\lambda)}(\zeta,\vec{k}), v_{nlm(\lambda)}(\zeta,\beta\theta\varphi). \tag{18}$$

See Ref. [20] for the exact definition of functions occurring in Eqs. (13) - (18).

The relativistic spinor orbitals satisfy the following orthogonality relations:

$$\int K^{\alpha s\dagger}_{nljm}(\zeta,\vec{x}) \overline{K}^{\alpha s}_{n'l'j'm'}(\zeta,\vec{x}) d\vec{x} = \delta_{nn'}\delta_{ll'}\delta_{jj'}\delta_{mm'} \tag{19}$$

$$\int K^{s\dagger}_{nljm}(\zeta,\vec{x}) K^{s}_{n'l'j'm'}(\zeta,\vec{x}) d\vec{x} = \frac{(n+n')!}{\left[(2n)!(2n')!\right]^{1/2}} \delta_{ll'}\delta_{jj'}\delta_{mm'}. \tag{20}$$

Using the relation $a^0_{ljm}(\lambda) = \delta_{jl}\delta_{mm_l}\delta_{\lambda 0}$ and formulae

$$H^0_{ljm}(\theta,\varphi) = \beta_{m_l} Y_{lm_l}(\theta,\varphi) \tag{21a}$$

$$\mathcal{H}^0_{ljm}(\theta,\varphi) = -i\beta_{m_l} Y_{lm_l}(\theta,\varphi) \tag{21b}$$

for the scalar particles it is easy to show that the relativistic spinor functions $K^{\alpha s}_{nljm}$, $\overline{K}^{\alpha s}_{nljm}$ and relativistic Slater spinor orbitals $K^{s}_{nljm}$ for particles with spin s=0 are reduced to the corresponding quantities for nonrelativistic complete basis sets in positon, momentum and four-dimensional spaces, i.e., $K^{\alpha s}_{nljm} \equiv k^{\alpha}_{nlm_l}$, $\overline{K}^{\alpha s}_{nljm} \equiv \overline{k}^{\alpha}_{nlm_l}$ and $K^{s}_{nljm} \equiv k_{nlm_l}$, where $s=0$, $j=l$, $t=0$, $m(\lambda) = m_l \delta_{\lambda 0}$ and $m = m_l$. Thus, the nonrelativistic and relativistic scalar particles can be also described by wave functions $K^{\alpha s}_{nljm}, \overline{K}^{\alpha s}_{nljm}$ and $K^{s}_{nljm}$ for s=t=0, $j=l$ and $m=m_l$, i.e.,

$$K^{\alpha 0}_{nljm} = \frac{1}{\sqrt{2}}\begin{bmatrix}1\\-i\end{bmatrix} k^{\alpha}_{nlm_l} \tag{22a}$$

$$\overline{K}^{\alpha 0}_{nljm} = \frac{1}{\sqrt{2}}\begin{bmatrix}1\\-i\end{bmatrix} \overline{k}^{\alpha}_{nlm_l} \tag{22b}$$

$$K^{0}_{nljm} = \frac{1}{\sqrt{2}}\begin{bmatrix}1\\-i\end{bmatrix} k_{nlm_l}. \tag{23}$$

The 6-, 10- and 4-, 8-component complete orthonormal basis sets of relativistic $\Psi^{\alpha s}$-ETSO through the nonrelativistic $\psi^{\alpha}$-ETO in position space for $s=1, s=2$ and $s=\frac{1}{2}, s=\frac{3}{2}$, respectively, are given in Tables 1, 2 and 3, 4.

## 4. Derivatives of $\Psi^{\alpha s}$-ETSO in position space

Now, we evaluate the derivatives of $\Psi^{\alpha s}$-ETSO with respect to Cartesian coordinates that can be used in the solution of reduced Dirac equations when the LCAO approach is employed. For this purpose we use the $\Psi^{\alpha s}$-ETSO in the following form:

for integral spin

$$\Psi^{\alpha s}_{nljm} = \frac{1}{\sqrt{2}} \begin{bmatrix} \phi^{\alpha s}_{nljm} \\ \tilde{\phi}^{\alpha s}_{\tilde{n}ljm} \end{bmatrix} = \frac{1}{\sqrt{2}} \begin{bmatrix} \phi^{\alpha s,0}_{nljm} \\ \phi^{\alpha s,1}_{nljm} \\ \vdots \\ \phi^{\alpha s,2s}_{nljm} \\ \tilde{\phi}^{\alpha s,2s}_{\tilde{n}ljm} \\ \vdots \\ \tilde{\phi}^{\alpha s,1}_{\tilde{n}ljm} \\ \tilde{\phi}^{\alpha s,0}_{\tilde{n}ljm} \end{bmatrix}, \qquad (24)$$

for half integral spin

$$\Psi^{\alpha s}_{nljm} = \frac{1}{\sqrt{2}} \begin{bmatrix} \varphi^{\alpha s}_{nljm} \\ \tilde{\varphi}^{\alpha s}_{\tilde{n}ljm} \end{bmatrix} = \frac{1}{\sqrt{2}} \begin{bmatrix} \varphi^{\alpha s,0}_{nljm} \\ \varphi^{\alpha s,2}_{nljm} \\ \vdots \\ \varphi^{\alpha s,2s-1}_{nljm} \\ \tilde{\varphi}^{\alpha s,2s-1}_{\tilde{n}ljm} \\ \vdots \\ \tilde{\varphi}^{\alpha s,2}_{\tilde{n}ljm} \\ \tilde{\varphi}^{\alpha s,0}_{\tilde{n}ljm} \end{bmatrix}, \qquad (25)$$

where $\phi^{\alpha s,\lambda}, \tilde{\phi}^{\alpha s,\lambda}$ and $\varphi^{\alpha s,\lambda}, \tilde{\varphi}^{\alpha s,\lambda}$ are the one- and two-component spinors, respectively, are defined by

$$\phi^{\alpha s,\lambda}_{nljm} = R^{\alpha}_{nl}(r) H^{s\lambda}_{ljm}(\theta,\varphi) \qquad (26a)$$

$$\tilde{\phi}^{\alpha s,\lambda}_{\tilde{n}ljm} = \tilde{R}^{\alpha}_{\tilde{n}l}(r) \mathcal{H}^{s\lambda}_{ljm}(\theta,\varphi) \qquad (26b)$$

$$\varphi^{\alpha s,\lambda}_{nljm} = R^{\alpha}_{nl}(r) Y^{s\lambda}_{ljm}(\theta,\varphi) \qquad (27a)$$

$$\tilde{\varphi}^{\alpha s,\lambda}_{\tilde{n}ljm} = \tilde{R}^{\alpha}_{\tilde{n}l}(r) \Upsilon^{s\lambda}_{ljm}(\theta,\varphi) \qquad (27b)$$

Here, $0 \leq \lambda(1) \leq 2s$ and $0 \leq \lambda(2) \leq 2s-1$ for integral and half-integral spin, respectively.

To obtain the derivatives of $\Psi^{\alpha s}$-ETSO we use the following relations:

$$\frac{\partial}{\partial z}(f\beta_m Y_{lm}) = \sum_{k=-1}^{1}{}' \left[\frac{df}{dr} + (\delta_{k,-1} - kl)\frac{f}{r}\right] b_k^{lm}\beta_m Y_{l+k,m} \qquad (28)$$

$$\left(\frac{\partial}{\partial x} - i\frac{\partial}{\partial y}\right)(f\beta_m Y_{lm}) = \sum_{k=-1}^{1}{}' \left[\frac{df}{dr} + (\delta_{k,-1} - kl)\frac{f}{r}\right] d_k^{lm}\beta_{m-1} Y_{l+k,m-1} \qquad (29)$$

$$\left(\frac{\partial}{\partial x} + i\frac{\partial}{\partial y}\right)(f\beta_m Y_{lm}) = \sum_{k=-1}^{1}{}' \left[\frac{df}{dr} + (\delta_{k,-1} - kl)\frac{f}{r}\right] c_k^{lm}\beta_{m+1} Y_{l+k,m+1}, \qquad (30)$$

where $f$ is any function of the radial distance $r$ and

$$b_k^{lm} = \left[(l+m+\delta_{k1})(l-m+\delta_{k1})/(2(l+1)+k)(2l+k)\right]^{1/2} \qquad (31)$$

$$d_k^{lm} = -k\left[(l-km+2\delta_{k1})(l-k(m-1))/(2(l+1)+k)(2l+k)\right]^{1/2} \qquad (32)$$

$$c_k^{lm} = k\left[(l+km+2\delta_{k1})(l+k(m+1))/(2(l+1)+k)(2l+k)\right]^{1/2} = -d_k^{l,-m}. \qquad (33)$$

The symbol $\sum{}'$ in Eqs. (28), (29) and (30) indicates that the summation is to be performed in steps of two. These formulae can be obtained by the use of method set out in Ref.[21].

Using Eqs. (28), (29) and (30) we obtain for the derivatives of two-component spinors of half-integral spin the following relations:

$$c(\vec{\sigma}\hat{\vec{p}})\varphi^{\alpha s\lambda} = c(\vec{\sigma}\hat{\vec{p}})\left[R_{nl}^\alpha Y_{ljm}^{s\lambda}\right] = \frac{c\hbar}{i}\sum_{k=-1}^{1}{}'\left[\frac{dR_{nl}^\alpha}{dr} + (\delta_{k,-1}-kl)\frac{R_{nl}^\alpha}{r}\right]\times$$
$$\times\begin{bmatrix} {}_kB_{ljm}^s(\lambda;\theta,\varphi) + {}_kD_{ljm}^s(\lambda+1;\theta,\varphi) \\ {}_kC_{ljm}^s(\lambda;\theta,\varphi) - {}_kB_{ljm}^s(\lambda+1;\theta,\varphi) \end{bmatrix} \qquad (34)$$

$$c(\vec{\sigma}\hat{\vec{p}})\tilde{\varphi}^{\alpha s\lambda} = c(\vec{\sigma}\hat{\vec{p}})\left[\tilde{R}_{\tilde{n}l}^\alpha \Upsilon_{ljm}^{s\lambda}\right] = -c\hbar\sum_{k=-1}^{1}{}'\left[\frac{d\tilde{R}_{\tilde{n}l}^\alpha}{dr} + (\delta_{k,-1}-kl)\frac{\tilde{R}_{\tilde{n}l}^\alpha}{r}\right]\times$$
$$\times\begin{bmatrix} {}_kB_{ljm}^s(\lambda+1;\theta,\varphi) + {}_kD_{ljm}^s(\lambda;\theta,\varphi) \\ {}_kC_{ljm}^s(\lambda+1;\theta,\varphi) - {}_kB_{ljm}^s(\lambda;\theta,\varphi) \end{bmatrix}, \qquad (35)$$

where,

$${}_kB_{ljm}^s(\lambda;\theta,\varphi) = a_{ljm}^s(\lambda) b_k^{lm(\lambda)} \beta_{m(\lambda)} Y_{l+k,m(\lambda)}(\theta,\varphi) \qquad (36)$$

$${}_kC_{ljm}^s(\lambda;\theta,\varphi) = a_{ljm}^s(\lambda) c_k^{lm(\lambda)} \beta_{m(\lambda)+1} Y_{l+k,m(\lambda)+1}(\theta,\varphi) \qquad (37)$$

$${}_kD_{ljm}^s(\lambda;\theta,\varphi) = a_{ljm}^s(\lambda) d_k^{lm(\lambda)} \beta_{m(\lambda)-1} Y_{l+k,m(\lambda)-1}(\theta,\varphi). \qquad (38)$$

The formulae presented in this work show that all of the 2(2s+1)-component relativistic basis spinor wave functions and Slater basis spinor orbitals are expressed through the sets of one- and two-component basis spinors. The radial parts of these basis spinors are determined from the corresponding nonrelativistic basis functions defined in position, momentum and four-dimensional spaces. Thus, the expansion and one-range addition theorems established in [20] for the nonrelativistic $k^{\alpha}_{nlm_l}$ and $k_{nlm_l}$ basis sets in position, momentum and four-dimensional spaces can be also used in the case of relativistic basis spinor functions $K^{\alpha s}_{nljm}$ and $K^{s}_{nljm}$. Accordingly, the electronic structure properties of arbitrary spin relativistic systems can be investigated with the help of corresponding nonrelativistic calculations.

## 5. Conclusions

The relativistic basis sets of spinor orbitals for the arbitrary spin particles in position, momentum and four-dimensional spaces are constructed from the product of complete sets of radial orbitals of nonrelativistic $\psi^{\alpha}-ETO$ corresponding to the spinor type tensor spherical harmonics.

The relativistic basis spinors obtained have the following properties:

(1) The relativistic spinors possess 2(2s+1) independent components.

(2) The 2(2s+1)-component spinors of integral spin are reduced to the sets of one-component spinors.

(3) The 2(2s+1)-component spinors of half-integral spin are reduced to the sets of two-component spinors.

(4) The relativistic spinors with s = 0 are reduced to the nonrelativistic complete sets of orbitals.

Thus, we have described the combined study of relativistic and nonrelativistic basis sets of exponential type orbitals for arbitrary spin and scalar particles in position, momentum and four-dimensional spaces. The method here presented for developing the theory of basis sets of $\Psi^{\alpha s}$-ETSO and $X^{s}$-STSO in particularly evident in the application to the solution of different problems of describing arbitrary spin within the framework of relativistic quantum chemistry.

Table 1. The exponential type spinor orbitals in position space for $s=1$, $1 \leq n \leq 3$, $0 \leq l \leq n-1$, $|l-s| \leq j \leq l+s$, $-j \leq m \leq j$, $t=2(j-l)$ and $\tilde{n}=n+|t|$

| n | l | j | t | $\tilde{n}$ | m | $\tilde{\Psi}^{\alpha 1}_{nljm}$ | | | |
|---|---|---|---|---|---|---|---|---|---|
| 1 | 0 | 1 | 2 | 3 | 1  | $\dfrac{\psi^\alpha_{100}}{\sqrt{2}}$ | 0 | 0 | $-\dfrac{i\psi^\alpha_{300}}{\sqrt{2}}$ |
|   |   |   |   |   | 0  | 0 | $\dfrac{\psi^\alpha_{100}}{\sqrt{2}}$ | 0 | $-\dfrac{i\psi^\alpha_{300}}{\sqrt{2}}$ |
|   |   |   |   |   | -1 | 0 | 0 | $\dfrac{\psi^\alpha_{100}}{\sqrt{2}}$ | 0 |
| 2 | 0 | 1 | 2 | 4 | 1  | $\dfrac{\psi^\alpha_{200}}{\sqrt{2}}$ | 0 | 0 | $-\dfrac{i\psi^\alpha_{400}}{\sqrt{2}}$ |
|   |   |   |   |   | 0  | 0 | $\dfrac{\psi^\alpha_{200}}{\sqrt{2}}$ | 0 | $-\dfrac{i\psi^\alpha_{400}}{\sqrt{2}}$ |
|   |   |   |   |   | -1 | 0 | 0 | $\dfrac{\psi^\alpha_{200}}{\sqrt{2}}$ | 0 |
|   | 1 | 1 | 0 | 2 | 1  | $-\dfrac{1}{2}\psi^\alpha_{210}$ | $\dfrac{1}{2}\psi^\alpha_{211}$ | 0 | $\dfrac{1}{2}i\psi^\alpha_{210}$ |
|   |   |   |   |   | 0  | $\dfrac{1}{2}\psi^\alpha_{21-1}$ | 0 | $-\dfrac{1}{2}i\psi^\alpha_{211}$ | 0 |
|   |   |   |   |   | -1 | 0 | $-\dfrac{1}{2}\psi^\alpha_{21-1}$ | $-\dfrac{1}{2}i\psi^\alpha_{210}$ | $-\dfrac{1}{2}i\psi^\alpha_{21-1}$ |
|   |   | 2 | 2 | 4 | 2  | $\dfrac{\psi^\alpha_{211}}{\sqrt{2}}$ | 0 | 0 | $-\dfrac{i\psi^\alpha_{411}}{\sqrt{2}}$ |
|   |   |   |   |   | 1  | $\dfrac{1}{2}\psi^\alpha_{210}$ | $\dfrac{1}{2}\psi^\alpha_{211}$ | 0 | $-\dfrac{1}{2}i\psi^\alpha_{410}$ |
|   |   |   |   |   | 0  | $\dfrac{\psi^\alpha_{21-1}}{2\sqrt{3}}$ | $\dfrac{\psi^\alpha_{210}}{\sqrt{3}}$ | $\dfrac{\psi^\alpha_{211}}{2\sqrt{3}}$ | $\dfrac{i\psi^\alpha_{41-1}}{2\sqrt{3}}$ |
|   |   |   |   |   | -1 | 0 | $-\dfrac{1}{2}\psi^\alpha_{21-1}$ | $\dfrac{1}{2}\psi^\alpha_{210}$ | 0 |

| 3 | 0 | 1 | 2 | 5 | -2 | 0 | 0 | 0 | 0 | $-\frac{i\psi^\alpha_{500}}{\sqrt{2}}$ | 0 |
| | | | | 5 | 1 | $\frac{\psi^\alpha_{300}}{\sqrt{2}}$ | 0 | 0 | $\frac{i\psi^\alpha_{41-1}}{\sqrt{2}}$ | 0 | $-\frac{i\psi^\alpha_{500}}{\sqrt{2}}$ |
| | | | | | 0 | 0 | $\frac{\psi^\alpha_{300}}{\sqrt{2}}$ | 0 | 0 | $\frac{i\psi^\alpha_{500}}{\sqrt{2}}$ | 0 |
| | | | | | -1 | 0 | 0 | $-\frac{\psi^\alpha_{300}}{\sqrt{2}}$ | $-\frac{i\psi^\alpha_{500}}{\sqrt{2}}$ | 0 | 0 |
| 1 | 1 | 1 | 0 | 3 | 1 | $-\frac{1}{2}\psi^\alpha_{310}$ | $\frac{1}{2}\psi^\alpha_{311}$ | 0 | 0 | $-\frac{1}{2}i\psi^\alpha_{311}$ | $\frac{1}{2}i\psi^\alpha_{310}$ |
| | | | | | 0 | $\frac{1}{2}\psi^\alpha_{31-1}$ | 0 | $\frac{1}{2}\psi^\alpha_{311}$ | $-\frac{1}{2}i\psi^\alpha_{311}$ | 0 | $-\frac{1}{2}i\psi^\alpha_{31-1}$ |
| | | | | | -1 | 0 | $\frac{1}{2}\psi^\alpha_{31-1}$ | $\frac{1}{2}\psi^\alpha_{310}$ | $-\frac{1}{2}i\psi^\alpha_{310}$ | $-\frac{1}{2}i\psi^\alpha_{31-1}$ | 0 |
| | 1 | 2 | 2 | 5 | 2 | $\frac{\psi^\alpha_{311}}{\sqrt{2}}$ | 0 | 0 | 0 | 0 | $-\frac{i\psi^\alpha_{511}}{\sqrt{2}}$ |
| | | | | | 1 | $\frac{1}{2}\psi^\alpha_{310}$ | $\frac{1}{2}\psi^\alpha_{311}$ | $\frac{\psi^\alpha_{311}}{2\sqrt{3}}$ | $\frac{i\psi^\alpha_{511}}{\sqrt{3}}$ | $-\frac{1}{2}i\psi^\alpha_{511}$ | $-\frac{1}{2}i\psi^\alpha_{510}$ |
| | | | | | 0 | $\frac{\psi^\alpha_{31-1}}{2\sqrt{3}}$ | $\frac{\psi^\alpha_{310}}{\sqrt{3}}$ | $-\frac{1}{2}i\psi^\alpha_{510}$ | $-\frac{1}{2}i\psi^\alpha_{510}$ | $\frac{i\psi^\alpha_{510}}{\sqrt{3}}$ | $\frac{i\psi^\alpha_{51-1}}{2\sqrt{3}}$ |
| | | | | | -1 | 0 | $-\frac{1}{2}\psi^\alpha_{31-1}$ | $-\frac{\psi^\alpha_{31-1}}{\sqrt{2}}$ | $\frac{i\psi^\alpha_{51-1}}{\sqrt{2}}$ | $\frac{1}{2}i\psi^\alpha_{51-1}$ | 0 |
| | | | | | -2 | 0 | 0 | $\sqrt{\frac{3}{10}}\psi^\alpha_{322}$ | $-\sqrt{\frac{3}{10}}i\psi^\alpha_{522}$ | 0 | 0 |
| | 2 | 1 | -2 | 5 | 1 | $\frac{\psi^\alpha_{320}}{2\sqrt{5}}$ | $-\frac{\psi^\alpha_{320}}{\sqrt{5}}$ | $\frac{1}{2}\sqrt{\frac{3}{5}}\psi^\alpha_{321}$ | $-\frac{1}{2}\sqrt{\frac{3}{5}}i\psi^\alpha_{521}$ | $\frac{i\psi^\alpha_{520}}{\sqrt{5}}$ | $-\frac{i\psi^\alpha_{520}}{2\sqrt{5}}$ |
| | | | | | 0 | $-\frac{1}{2}\sqrt{\frac{3}{5}}\psi^\alpha_{32-1}$ | $\frac{1}{2}\sqrt{\frac{3}{5}}\psi^\alpha_{321}$ | $\frac{1}{2}\sqrt{\frac{3}{5}}i\psi^\alpha_{521}$ | 0 | $\frac{1}{2}\sqrt{\frac{3}{5}}i\psi^\alpha_{52-1}$ |

| | | | -1 | $-\sqrt{\frac{3}{10}}\psi^\alpha_{32-2}$ | $\frac{1}{2}\sqrt{\frac{3}{5}}\psi^\alpha_{32-1}$ | $\frac{\psi^\alpha_{320}}{2\sqrt{5}}$ | $-\frac{i\psi^\alpha_{520}}{2\sqrt{5}}$ | $-\frac{1}{2}\sqrt{\frac{3}{5}}i\psi^\alpha_{52-1}$ | $-\sqrt{\frac{3}{10}}i\psi^\alpha_{52-2}$ |
|---|---|---|---|---|---|---|---|---|---|
| | 3 | 0 | 2 | $-\frac{\psi^\alpha_{321}}{\sqrt{6}}$ | $\frac{\psi^\alpha_{322}}{\sqrt{3}}$ | 0 | 0 | $-\frac{i\psi^\alpha_{322}}{\sqrt{3}}$ | $\frac{i\psi^\alpha_{321}}{\sqrt{6}}$ |
| 2 | | | 1 | $-\frac{1}{2}\psi^\alpha_{320}$ | $\frac{\psi^\alpha_{321}}{2\sqrt{3}}$ | $\frac{\psi^\alpha_{322}}{\sqrt{6}}$ | $-\frac{i\psi^\alpha_{322}}{\sqrt{6}}$ | $-\frac{i\psi^\alpha_{321}}{2\sqrt{3}}$ | $\frac{1}{2}i\psi^\alpha_{320}$ |
| | | | 0 | $\frac{1}{2}\psi^\alpha_{32-1}$ | 0 | $-\frac{1}{2}\psi^\alpha_{321}$ | $-\frac{1}{2}i\psi^\alpha_{321}$ | 0 | $-\frac{1}{2}i\psi^\alpha_{32-1}$ |
| | | | -1 | $-\frac{\psi^\alpha_{32-2}}{\sqrt{6}}$ | $\frac{\psi^\alpha_{32-1}}{2\sqrt{3}}$ | $\frac{1}{2}\psi^\alpha_{320}$ | $-\frac{1}{2}i\psi^\alpha_{320}$ | $-\frac{i\psi^\alpha_{32-1}}{2\sqrt{3}}$ | $\frac{i\psi^\alpha_{32-2}}{\sqrt{6}}$ |
| | | | -2 | 0 | $-\frac{\psi^\alpha_{32-2}}{\sqrt{3}}$ | $\frac{\psi^\alpha_{32-1}}{\sqrt{6}}$ | $\frac{i\psi^\alpha_{32-1}}{\sqrt{6}}$ | $\frac{i\psi^\alpha_{32-2}}{\sqrt{3}}$ | 0 |
| | 5 | 2 | 3 | $\frac{\psi^\alpha_{322}}{\sqrt{2}}$ | 0 | 0 | 0 | 0 | $\frac{i\psi^\alpha_{522}}{\sqrt{2}}$ |
| 3 | | | 2 | $\frac{\psi^\alpha_{321}}{\sqrt{3}}$ | $\frac{\psi^\alpha_{322}}{\sqrt{6}}$ | $\frac{\psi^\alpha_{522}}{\sqrt{30}}$ | $\frac{i\psi^\alpha_{522}}{\sqrt{30}}$ | $\frac{i\psi^\alpha_{521}}{\sqrt{6}}$ | $-\frac{i\psi^\alpha_{521}}{\sqrt{3}}$ |
| | | | 1 | $\frac{A\psi^\alpha_{320}}{\sqrt{5}}$ | $\frac{2\psi^\alpha_{321}}{\sqrt{15}}$ | $\frac{\psi^\alpha_{321}}{\sqrt{10}}$ | $-\frac{i\psi^\alpha_{521}}{\sqrt{10}}$ | $\frac{2i\psi^\alpha_{521}}{\sqrt{15}}$ | $-\frac{i\psi^\alpha_{520}}{\sqrt{5}}$ |
| | | | 0 | $-\frac{\psi^\alpha_{32-1}}{\sqrt{10}}$ | $\sqrt{\frac{3}{10}}\psi^\alpha_{320}$ | $\frac{\psi^\alpha_{320}}{\sqrt{10}}$ | $-\frac{i\psi^\alpha_{520}}{\sqrt{10}}$ | $-\sqrt{\frac{3}{10}}i\psi^\alpha_{520}$ | $\frac{i\psi^\alpha_{52-1}}{\sqrt{10}}$ |
| | | | -1 | $\frac{\psi^\alpha_{32-2}}{\sqrt{30}}$ | $\frac{2\psi^\alpha_{32-1}}{\sqrt{15}}$ | $\frac{\psi^\alpha_{320}}{\sqrt{5}}$ | $-\frac{i\psi^\alpha_{520}}{\sqrt{5}}$ | $\frac{2i\psi^\alpha_{52-1}}{\sqrt{15}}$ | $-\frac{i\psi^\alpha_{521}}{\sqrt{3}}$ |
| | | | -2 | 0 | $-\frac{\psi^\alpha_{32-2}}{\sqrt{6}}$ | $\frac{\psi^\alpha_{32-1}}{\sqrt{3}}$ | $-\frac{i\psi^\alpha_{52-1}}{\sqrt{3}}$ | $\frac{i\psi^\alpha_{52-2}}{\sqrt{6}}$ | 0 |
| | | | -3 | 0 | 0 | $\frac{\psi^\alpha_{32-2}}{\sqrt{2}}$ | $\frac{i\psi^\alpha_{52-2}}{\sqrt{2}}$ | 0 | 0 |

Table 2. The exponential type spinor orbitals in position space for $s=2$, $1 \leq n \leq 3$, $0 \leq l \leq n-1$, $|l-s| \leq j \leq l+s$, $-j \leq m \leq j$, $t = 2(j-l)$ and $\tilde{n} = n + |t|$

| n | l | j | t | $\tilde{n}$ | m | | | | | $\tilde{\Psi}^{\alpha 2}_{nljm}$ | | | | | |
|---|---|---|---|---|---|---|---|---|---|---|---|---|---|---|---|
| 1 | 0 | 2 | 4 | 5 | 2 | $\frac{\psi^\alpha_{100}}{\sqrt{2}}$ | 0 | 0 | 0 | 0 | 0 | 0 | 0 | 0 | $-\frac{i\psi^\alpha_{500}}{\sqrt{2}}$ |
| | | | | | 1 | 0 | $\frac{\psi^\alpha_{100}}{\sqrt{2}}$ | 0 | 0 | 0 | 0 | 0 | 0 | $-\frac{i\psi^\alpha_{500}}{\sqrt{2}}$ | 0 |
| | | | | | 0 | 0 | 0 | $\frac{\psi^\alpha_{100}}{\sqrt{2}}$ | 0 | 0 | 0 | 0 | $-\frac{i\psi^\alpha_{500}}{\sqrt{2}}$ | 0 | 0 |
| | | | | | -1 | 0 | 0 | 0 | $\frac{\psi^\alpha_{100}}{\sqrt{2}}$ | 0 | 0 | $-\frac{i\psi^\alpha_{500}}{\sqrt{2}}$ | 0 | 0 | 0 |
| | | | | | -2 | 0 | 0 | 0 | 0 | $\frac{\psi^\alpha_{100}}{\sqrt{2}}$ | $-\frac{i\psi^\alpha_{500}}{\sqrt{2}}$ | 0 | 0 | 0 | 0 |
| 2 | 0 | 2 | 4 | 6 | 2 | $\frac{\psi^\alpha_{200}}{\sqrt{2}}$ | 0 | 0 | 0 | 0 | 0 | 0 | 0 | 0 | $-\frac{i\psi^\alpha_{600}}{\sqrt{2}}$ |
| | | | | | 1 | 0 | $\frac{\psi^\alpha_{200}}{\sqrt{2}}$ | 0 | 0 | 0 | 0 | 0 | 0 | $-\frac{i\psi^\alpha_{600}}{\sqrt{2}}$ | 0 |
| | | | | | 0 | 0 | 0 | $\frac{\psi^\alpha_{200}}{\sqrt{2}}$ | 0 | 0 | 0 | 0 | $-\frac{i\psi^\alpha_{600}}{\sqrt{2}}$ | 0 | 0 |
| | | | | | -1 | 0 | 0 | 0 | $\frac{\psi^\alpha_{200}}{\sqrt{2}}$ | 0 | 0 | $-\frac{i\psi^\alpha_{600}}{\sqrt{2}}$ | 0 | 0 | 0 |
| | | | | | -2 | 0 | 0 | 0 | 0 | $\frac{\psi^\alpha_{200}}{\sqrt{2}}$ | $-\frac{i\psi^\alpha_{600}}{\sqrt{2}}$ | 0 | 0 | 0 | 0 |

| | | | | | | | | | | | | |
|---|---|---|---|---|---|---|---|---|---|---|---|---|
| 1 | | | | | | | | | | | | |
| | 2 | 2 | | | | | | | | | | |
| | | | 2 | $-\frac{\psi^\alpha_{210}}{\sqrt{3}}$ | $\frac{\psi^\alpha_{211}}{\sqrt{6}}$ | 0 | 0 | 0 | 0 | 0 | $-\frac{i\psi^\alpha_{411}}{\sqrt{6}}$ | $\frac{i\psi^\alpha_{410}}{\sqrt{3}}$ |
| | | | 1 | $-\frac{\psi^\alpha_{21-1}}{\sqrt{6}}$ | $-\frac{\psi^\alpha_{210}}{2\sqrt{3}}$ | $\frac{1}{2}\psi^\alpha_{211}$ | 0 | 0 | 0 | $-\frac{1}{2}i\psi^\alpha_{411}$ | $\frac{i\psi^\alpha_{410}}{2\sqrt{3}}$ | $-\frac{i\psi^\alpha_{41-1}}{\sqrt{6}}$ |
| | | | 0 | 0 | $\frac{1}{2}\psi^\alpha_{21-1}$ | 0 | $\frac{1}{2}\psi^\alpha_{2,1,1}$ | 0 | 0 | 0 | 0 | $-\frac{1}{2}i\psi^\alpha_{41-1}$ |
| | | | -1 | 0 | 0 | $\frac{1}{2}\psi^\alpha_{21-1}$ | $\frac{\psi^\alpha_{210}}{2\sqrt{3}}$ | $\frac{\psi^\alpha_{211}}{\sqrt{6}}$ | 0 | $-\frac{1}{2}i\psi^\alpha_{411}$ | 0 | 0 |
| | | | -2 | 0 | 0 | 0 | $-\frac{\psi^\alpha_{21-1}}{\sqrt{6}}$ | $-\frac{\psi^\alpha_{210}}{\sqrt{3}}$ | 0 | 0 | 0 | 0 |
| | | 4 | | | | | | | | | | |
| | 2 | | | | | | | | | | | |
| | | | 3 | $\frac{\psi^\alpha_{211}}{\sqrt{2}}$ | $\frac{\psi^\alpha_{211}}{\sqrt{3}}$ | $\frac{\psi^\alpha_{210}}{\sqrt{5}}$ | 0 | 0 | $-\frac{i\psi^\alpha_{611}}{\sqrt{2}}$ | $\frac{i\psi^\alpha_{611}}{\sqrt{3}}$ | $-\frac{i\psi^\alpha_{611}}{\sqrt{2}}$ |
| | | 6 | | | | | | | | | | |
| | | | 2 | $\frac{\psi^\alpha_{210}}{\sqrt{6}}$ | $\frac{2\psi^\alpha_{210}}{\sqrt{15}}$ | $\sqrt{\frac{3}{10}}\psi^\alpha_{210}$ | $\frac{\psi^\alpha_{211}}{\sqrt{10}}$ | $\frac{\psi^\alpha_{211}}{\sqrt{30}}$ | $\frac{i\psi^\alpha_{611}}{\sqrt{3}}$ | $\frac{2i\psi^\alpha_{610}}{\sqrt{15}}$ | $-\frac{i\psi^\alpha_{610}}{\sqrt{6}}$ |
| | | | 1 | $-\frac{\psi^\alpha_{21-1}}{\sqrt{30}}$ | $-\frac{\psi^\alpha_{21-1}}{\sqrt{10}}$ | $-\frac{\psi^\alpha_{210}}{\sqrt{5}}$ | $\frac{2\psi^\alpha_{210}}{\sqrt{15}}$ | $\frac{\psi^\alpha_{210}}{\sqrt{6}}$ | $\frac{i\psi^\alpha_{61-1}}{\sqrt{30}}$ | $\frac{i\psi^\alpha_{611}}{\sqrt{5}}$ | $\frac{i\psi^\alpha_{61-1}}{\sqrt{30}}$ |
| | | | 0 | 0 | 0 | $-\sqrt{\frac{3}{10}}\psi^\alpha_{210}$ | $-\frac{\psi^\alpha_{21-1}}{\sqrt{10}}$ | $\frac{\psi^\alpha_{211}}{\sqrt{30}}$ | 0 | $-\sqrt{\frac{3}{10}}i\psi^\alpha_{610}$ | $\frac{i\psi^\alpha_{61-1}}{\sqrt{10}}$ | 0 |
| | | | -1 | 0 | 0 | 0 | $\frac{2i\psi^\alpha_{610}}{\sqrt{15}}$ | $\frac{i\psi^\alpha_{611}}{\sqrt{5}}$ | 0 | 0 | 0 | 0 |
| | | | -2 | 0 | 0 | 0 | $\frac{i\psi^\alpha_{61-1}}{\sqrt{3}}$ | $\frac{i\psi^\alpha_{61-1}}{\sqrt{5}}$ | 0 | 0 | 0 | 0 |

| n | l | m | | | | | | | | | | | |
|---|---|---|---|---|---|---|---|---|---|---|---|---|---|
| 3 | 0 | 2 | 4 | 7 | -3 | 0 | 0 | 0 | 0 | 0 | 0 | 0 | 0 |
|   |   |   |   |   | 2 | $\frac{\psi^\alpha_{300}}{\sqrt{2}}$ | 0 | 0 | 0 | 0 | 0 | 0 | $-\frac{i\psi^\alpha_{700}}{\sqrt{2}}$ | 0 |
|   |   |   |   |   | 1 | 0 | $\frac{\psi^\alpha_{300}}{\sqrt{2}}$ | 0 | 0 | 0 | 0 | $-\frac{i\psi^\alpha_{700}}{\sqrt{2}}$ | 0 | 0 |
|   |   |   |   |   | 0 | 0 | 0 | $\frac{\psi^\alpha_{300}}{\sqrt{2}}$ | 0 | 0 | $-\frac{i\psi^\alpha_{700}}{\sqrt{2}}$ | 0 | 0 | 0 |
|   |   |   |   |   | -1 | 0 | 0 | 0 | $\frac{\psi^\alpha_{300}}{\sqrt{2}}$ | $-\frac{i\psi^\alpha_{700}}{\sqrt{2}}$ | 0 | 0 | 0 | 0 |
|   |   |   |   |   | -2 | 0 | 0 | 0 | 0 | $\frac{i\psi^\alpha_{61-1}}{\sqrt{2}}$ | 0 | 0 | 0 | 0 |
| 1 | 2 | 2 | 2 | 5 | 2 | $-\frac{\psi^\alpha_{310}}{\sqrt{3}}$ | $\frac{\psi^\alpha_{311}}{\sqrt{6}}$ | 0 | 0 | $-\frac{\psi^\alpha_{21-1}}{\sqrt{2}}$ | $\frac{i\psi^\alpha_{511}}{\sqrt{6}}$ | $-\frac{i\psi^\alpha_{511}}{2}$ | $-\frac{i\psi^\alpha_{511}}{\sqrt{6}}$ | $\frac{i\psi^\alpha_{510}}{\sqrt{3}}$ |
|   |   |   |   |   | 1 | $-\frac{\psi^\alpha_{31-1}}{\sqrt{6}}$ | $-\frac{\psi^\alpha_{310}}{2\sqrt{3}}$ | $\frac{1}{2}\psi^\alpha_{311}$ | 0 | 0 | $-\frac{1}{2}i\psi^\alpha_{511}$ | $\frac{i\psi^\alpha_{510}}{2\sqrt{3}}$ | $\frac{i\psi^\alpha_{510}}{2\sqrt{3}}$ | $-\frac{i\psi^\alpha_{51-1}}{\sqrt{6}}$ |
|   |   |   |   |   | 0 | 0 | $\frac{1}{2}\psi^\alpha_{31-1}$ | 0 | $\frac{1}{2}\psi^\alpha_{311}$ | 0 | 0 | $-\frac{1}{2}i\psi^\alpha_{511}$ | $-\frac{1}{2}i\psi^\alpha_{51-1}$ | 0 |
|   |   |   |   |   | -1 | 0 | 0 | $\frac{1}{2}\psi^\alpha_{31-1}$ | $\frac{\psi^\alpha_{310}}{2\sqrt{3}}$ | $\frac{\psi^\alpha_{311}}{\sqrt{6}}$ | $-\frac{1}{2}i\psi^\alpha_{51-1}$ | 0 | 0 | 0 |

| | | | | | | | | | |
|---|---|---|---|---|---|---|---|---|---|
| -2 | 3 | 2 | 1 | 0 | -1 | -2 | -3 | 2 | 1 |
| 0 | $\dfrac{\psi^\alpha_{311}}{\sqrt{2}}$ | $\dfrac{\psi^\alpha_{310}}{\sqrt{6}}$ | $-\dfrac{\psi^\alpha_{31-1}}{\sqrt{30}}$ | 0 | 0 | 0 | 0 | $\dfrac{\psi^\alpha_{320}}{\sqrt{7}}$ | $-\sqrt{\dfrac{3}{14}}\dfrac{\psi^\alpha_{32-1}}{}$ |
| 0 | 0 | $\dfrac{\psi^\alpha_{311}}{\sqrt{3}}$ | $\dfrac{2\psi^\alpha_{310}}{\sqrt{15}}$ | $-\dfrac{\psi^\alpha_{31-1}}{\sqrt{10}}$ | 0 | 0 | 0 | $-\sqrt{\dfrac{3}{14}}\psi^\alpha_{321}$ | $\dfrac{\psi^\alpha_{320}}{2\sqrt{7}}$ |
| 0 | 0 | 0 | $\dfrac{\psi^\alpha_{311}}{\sqrt{5}}$ | $-\sqrt{\dfrac{3}{10}}\psi^\alpha_{310}$ | $\dfrac{\psi^\alpha_{31-1}}{\sqrt{5}}$ | 0 | 0 | $\dfrac{\psi^\alpha_{322}}{\sqrt{7}}$ | $-\dfrac{\psi^\alpha_{321}}{2\sqrt{7}}$ |
| $\dfrac{\psi^\alpha_{31-1}}{\sqrt{6}}$ | 0 | 0 | 0 | $\dfrac{\psi^\alpha_{311}}{\sqrt{10}}$ | $\dfrac{2\psi^\alpha_{310}}{\sqrt{15}}$ | $-\dfrac{\psi^\alpha_{31-1}}{\sqrt{3}}$ | 0 | 0 | $\sqrt{\dfrac{3}{14}}\psi^\alpha_{322}$ |
| $\dfrac{\psi^\alpha_{310}}{\sqrt{3}}$ | 0 | 0 | 0 | 0 | $\dfrac{\psi^\alpha_{311}}{\sqrt{30}}$ | $\dfrac{\psi^\alpha_{310}}{\sqrt{6}}$ | $-\dfrac{\psi^\alpha_{31-1}}{\sqrt{2}}$ | 0 | 0 |
| $-\dfrac{i\psi^\alpha_{51-1}}{\sqrt{6}}$ | 0 | 0 | 0 | $-\dfrac{i\psi^\alpha_{711}}{\sqrt{10}}$ | $-\dfrac{2i\psi^\alpha_{710}}{\sqrt{15}}$ | $\dfrac{i\psi^\alpha_{71-1}}{\sqrt{3}}$ | 0 | 0 | $-\sqrt{\dfrac{3}{14}}i\psi^\alpha_{322}$ |
| | 0 | 0 | $-\dfrac{i\psi^\alpha_{711}}{\sqrt{5}}$ | $-\sqrt{\dfrac{3}{10}}i\psi^\alpha_{710}$ | $\dfrac{i\psi^\alpha_{71-1}}{\sqrt{5}}$ | 0 | 0 | $-\dfrac{i\psi^\alpha_{322}}{\sqrt{7}}$ | $\dfrac{i\psi^\alpha_{321}}{2\sqrt{7}}$ |
| | 0 | 0 | $-\dfrac{i\psi^\alpha_{711}}{\sqrt{3}}$ | $-\dfrac{2i\psi^\alpha_{710}}{\sqrt{15}}$ | $\dfrac{i\psi^\alpha_{71-1}}{\sqrt{10}}$ | 0 | 0 | $\sqrt{\dfrac{3}{14}}i\psi^\alpha_{321}$ | $\dfrac{i\psi^\alpha_{320}}{2\sqrt{7}}$ |
| | $-\dfrac{i\psi^\alpha_{711}}{\sqrt{2}}$ | $-\dfrac{i\psi^\alpha_{710}}{\sqrt{6}}$ | $\dfrac{i\psi^\alpha_{71-1}}{\sqrt{30}}$ | 0 | 0 | 0 | 0 | $-\dfrac{i\psi^\alpha_{320}}{\sqrt{7}}$ | $\sqrt{\dfrac{3}{14}}i\psi^\alpha_{32-1}$ |

| 7 | | | | | | | | 3 | |
| 4 | | | | | | | | 0 | |
| 3 | | | | | | | | 2 | |
| | | | | | | | | 2 | |
| 0 | $\frac{\psi^\alpha_{32-2}}{\sqrt{7}}$ | | $-\frac{\psi^\alpha_{320}}{\sqrt{7}}$ | $\frac{\psi^\alpha_{321}}{2\sqrt{7}}$ | $\frac{\psi^\alpha_{322}}{\sqrt{7}}$ | $-\frac{i\psi^\alpha_{322}}{\sqrt{7}}$ | $\frac{i\psi^\alpha_{321}}{2\sqrt{7}}$ | $\frac{i\psi^\alpha_{320}}{\sqrt{7}}$ | $\frac{i\psi^\alpha_{32-1}}{2\sqrt{7}}$ | $-\frac{i\psi^\alpha_{32-2}}{\sqrt{7}}$ |
| -1 | 0 | $-\frac{\psi^\alpha_{32-1}}{2\sqrt{7}}$ | $-\frac{\psi^\alpha_{32-1}}{2\sqrt{7}}$ | $-\frac{\psi^\alpha_{320}}{2\sqrt{7}}$ | $\sqrt{\frac{3}{14}}\psi^\alpha_{321}$ | $-\sqrt{\frac{3}{14}}i\psi^\alpha_{321}$ | $\frac{i\psi^\alpha_{320}}{2\sqrt{7}}$ | $-\frac{i\psi^\alpha_{32-1}}{2\sqrt{7}}$ | $-\sqrt{\frac{3}{14}}i\psi^\alpha_{32-2}$ | 0 |
| -2 | 0 | $\sqrt{\frac{3}{14}}\psi^\alpha_{32-2}$ | $\frac{\psi^\alpha_{32-2}}{\sqrt{7}}$ | $\frac{3}{\sqrt{14}}\psi^\alpha_{32-1}$ | $\frac{\psi^\alpha_{320}}{\sqrt{7}}$ | $-\frac{i\psi^\alpha_{320}}{\sqrt{7}}$ | $-\sqrt{\frac{3}{14}}i\psi^\alpha_{32-1}$ | $-\frac{i\psi^\alpha_{32-2}}{\sqrt{7}}$ | 0 | 0 |
| 3 | $-\frac{1}{2}\psi^\alpha_{321}$ | $\frac{1}{2}\psi^\alpha_{322}$ | 0 | 0 | 0 | 0 | 0 | $-\frac{1}{2}i\psi^\alpha_{522}$ | $-\frac{1}{2}i\psi^\alpha_{522}$ | $\frac{1}{2}i\psi^\alpha_{521}$ |
| 2 | $-\frac{1}{2}\psi^\alpha_{320}$ | 0 | $\frac{1}{2}\psi^\alpha_{322}$ | 0 | 0 | $\frac{\psi^\alpha_{522}}{2\sqrt{5}}$ | $-\frac{1}{2}i\psi^\alpha_{522}$ | 0 | $-\frac{i\psi^\alpha_{522}}{2\sqrt{5}}$ | $\frac{1}{2}i\psi^\alpha_{520}$ |
| 1 | $\frac{1}{2}\sqrt{\frac{3}{5}}\psi^\alpha_{32-1}$ | $-\frac{\psi^\alpha_{320}}{\sqrt{10}}$ | $\frac{\psi^\alpha_{321}}{\sqrt{10}}$ | $\frac{1}{2}\sqrt{\frac{3}{5}}\psi^\alpha_{322}$ | $\frac{1}{2}\sqrt{\frac{3}{5}}\psi^\alpha_{321}$ | $\frac{\psi^\alpha_{522}}{2\sqrt{5}}$ | $-\frac{\psi^\alpha_{521}}{\sqrt{5}}$ | $-\frac{1}{2}i\psi^\alpha_{522}$ | 0 | $-\frac{1}{2}\sqrt{\frac{3}{5}}i\psi^\alpha_{52-1}$ |
| 0 | $-\frac{\psi^\alpha_{32-2}}{2\sqrt{5}}$ | $\frac{\psi^\alpha_{32-1}}{\sqrt{5}}$ | $\frac{\psi^\alpha_{320}}{\sqrt{10}}$ | $\frac{\psi^\alpha_{321}}{\sqrt{5}}$ | $\frac{\psi^\alpha_{322}}{2\sqrt{5}}$ | $\frac{i\psi^\alpha_{521}}{\sqrt{5}}$ | $\frac{i\psi^\alpha_{520}}{\sqrt{10}}$ | $\frac{i\psi^\alpha_{521}}{\sqrt{5}}$ | $-\frac{\psi^\alpha_{52-2}}{\sqrt{5}}$ | $-\frac{3}{2}\sqrt{\frac{3}{5}}i\psi^\alpha_{52-1}$ |
| -1 | 0 | 0 | $-\frac{1}{2}\psi^\alpha_{32-2}$ | $\frac{1}{2}\psi^\alpha_{32-1}$ | $\frac{1}{2}\psi^\alpha_{320}$ | $-\frac{1}{2}i\psi^\alpha_{52-1}$ | 0 | $\frac{1}{2}i\psi^\alpha_{520}$ | $\frac{i\psi^\alpha_{52-2}}{\sqrt{5}}$ | $\frac{i\psi^\alpha_{52-2}}{2\sqrt{5}}$ |
| -2 | 0 | 0 | 0 | $-\frac{1}{2}\psi^\alpha_{32-2}$ | $-\frac{1}{2}\psi^\alpha_{32-1}$ | $\frac{1}{2}i\psi^\alpha_{52-2}$ | 0 | 0 | $\frac{1}{2}i\psi^\alpha_{52-2}$ | 0 |
| -3 | 0 | 0 | 0 | 0 | 0 | 0 | 0 | 0 | 0 | 0 |

| | 4 | 3 | 2 | 1 | 0 | -1 | -2 | -3 | -4 |
|---|---|---|---|---|---|---|---|---|---|
| | $\dfrac{\psi^\alpha_{322}}{\sqrt{2}}$ | 0 | 0 | 0 | 0 | 0 | 0 | 0 | $-\dfrac{i\psi^\alpha_{722}}{\sqrt{2}}$ |
| | $\dfrac{1}{2}\psi^\alpha_{321}$ | $\dfrac{1}{2}\psi^\alpha_{322}$ | 0 | 0 | 0 | 0 | 0 | $-\dfrac{1}{2}i\psi^\alpha_{722}$ | $-\dfrac{1}{2}i\psi^\alpha_{721}$ |
| | $\dfrac{1}{2}\sqrt{\dfrac{3}{7}}\psi^\alpha_{320}$ | $\sqrt{\dfrac{2}{7}}\psi^\alpha_{321}$ | $\dfrac{1}{2}\sqrt{\dfrac{3}{7}}\psi^\alpha_{322}$ | 0 | 0 | 0 | $-\dfrac{1}{2}\sqrt{\dfrac{3}{7}}i\psi^\alpha_{722}$ | $-\sqrt{\dfrac{2}{7}}i\psi^\alpha_{721}$ | $-\dfrac{1}{2}\sqrt{\dfrac{3}{7}}i\psi^\alpha_{720}$ |
| | $-\dfrac{\psi^\alpha_{32-1}}{2\sqrt{7}}$ | $\sqrt{\dfrac{3}{14}}\psi^\alpha_{320}$ | $\dfrac{3}{\sqrt{35}}\psi^\alpha_{321}$ | $\dfrac{\psi^\alpha_{322}}{2\sqrt{7}}$ | 0 | $-\dfrac{i\psi^\alpha_{722}}{2\sqrt{35}}$ | $-\sqrt{\dfrac{3}{14}}i\psi^\alpha_{721}$ | $-\dfrac{3i\psi^\alpha_{720}}{\sqrt{35}}$ | $\dfrac{i\psi^\alpha_{72-1}}{2\sqrt{7}}$ |
| | $\dfrac{\psi^\alpha_{32-2}}{2\sqrt{35}}$ | $\dfrac{2\psi^\alpha_{32-1}}{\sqrt{35}}$ | $\dfrac{3\psi^\alpha_{320}}{\sqrt{35}}$ | $\dfrac{2\psi^\alpha_{321}}{\sqrt{35}}$ | $\dfrac{\psi^\alpha_{322}}{2\sqrt{35}}$ | $\dfrac{i\psi^\alpha_{722}}{2\sqrt{35}}$ | $\dfrac{2i\psi^\alpha_{721}}{\sqrt{35}}$ | $-\dfrac{3i\psi^\alpha_{720}}{\sqrt{35}}$ | $-\dfrac{i\psi^\alpha_{72-2}}{2\sqrt{7}}$ |
| | 0 | $\dfrac{\psi^\alpha_{32-2}}{2\sqrt{7}}$ | $-\sqrt{\dfrac{3}{14}}\psi^\alpha_{32-1}$ | $-\sqrt{\dfrac{2}{7}}\psi^\alpha_{32-0}$ | $\dfrac{\psi^\alpha_{321}}{2\sqrt{7}}$ | $-\dfrac{1}{2}\sqrt{\dfrac{3}{7}}i\psi^\alpha_{722}$ | $-\dfrac{3}{\sqrt{35}}i\psi^\alpha_{721}$ | $-\sqrt{\dfrac{3}{14}}i\psi^\alpha_{720}$ | $\dfrac{i\psi^\alpha_{72-1}}{2\sqrt{7}}$ |
| | 0 | 0 | $\dfrac{1}{2}\sqrt{\dfrac{3}{7}}\psi^\alpha_{32-2}$ | $\dfrac{3}{\sqrt{35}}\psi^\alpha_{32-1}$ | $\dfrac{\psi^\alpha_{320}}{\sqrt{35}}$ | $-\dfrac{1}{2}\sqrt{\dfrac{3}{7}}i\psi^\alpha_{721}$ | $-\dfrac{3}{\sqrt{35}}i\psi^\alpha_{720}$ | $\sqrt{\dfrac{3}{14}}i\psi^\alpha_{72-1}$ | $-\dfrac{1}{2}\sqrt{\dfrac{3}{7}}i\psi^\alpha_{72-2}$ |
| | 0 | 0 | 0 | $\dfrac{1}{2}\psi^\alpha_{32-2}$ | $\dfrac{2\psi^\alpha_{32-1}}{\sqrt{35}}$ | $-\dfrac{1}{2}i\psi^\alpha_{722}$ | $-\sqrt{\dfrac{2}{7}}i\psi^\alpha_{721}$ | $-\sqrt{\dfrac{3}{14}}i\psi^\alpha_{720}$ | $\dfrac{2i\psi^\alpha_{72-1}}{\sqrt{35}}$ |
| | 0 | 0 | 0 | 0 | $\dfrac{\psi^\alpha_{32-2}}{\sqrt{2}}$ | 0 | $-\dfrac{1}{2}i\psi^\alpha_{722}$ | $-\dfrac{1}{2}i\psi^\alpha_{721}$ | $\dfrac{i\psi^\alpha_{72-2}}{\sqrt{7}}$ |

al type spinor orbitals in position space for $s = \dfrac{1}{2}$, $1 \leq n \leq 3$, $0 \leq l \leq n-1$, $|l-s| \leq j \leq l+s$, $-j \leq m \leq j$, $t = 2(j-l)$ and $\tilde{n} = n + |t|$

| n | $l$ | j | t | $\tilde{n}$ | m | $\tilde{\Psi}_{nljm}^{\alpha 1/2}$ | | | |
|---|---|---|---|---|---|---|---|---|---|
| 1 | 0 | 1/2 | 1 | 2 | 1/2 | $\dfrac{\psi_{100}^{\alpha}}{\sqrt{2}}$ | 0 | 0 | $-\dfrac{i\psi_{200}^{\alpha}}{\sqrt{2}}$ |
| | | | | | -1/2 | 0 | $\dfrac{\psi_{100}^{\alpha}}{\sqrt{2}}$ | $-\dfrac{i\psi_{200}^{\alpha}}{\sqrt{2}}$ | 0 |
| 2 | 0 | 1/2 | 1 | 3 | 1/2 | $\dfrac{\psi_{200}^{\alpha}}{\sqrt{2}}$ | 0 | 0 | $-\dfrac{i\psi_{300}^{\alpha}}{\sqrt{2}}$ |
| | | | | | -1/2 | 0 | $\dfrac{\psi_{200}^{\alpha}}{\sqrt{2}}$ | $-\dfrac{i\psi_{300}^{\alpha}}{\sqrt{2}}$ | 0 |
| | 1 | 1/2 | -1 | 3 | 1/2 | $-\dfrac{\psi_{210}^{\alpha}}{\sqrt{6}}$ | $\dfrac{\psi_{211}^{\alpha}}{\sqrt{3}}$ | $-\dfrac{i\psi_{311}^{\alpha}}{\sqrt{3}}$ | $\dfrac{i\psi_{310}^{\alpha}}{\sqrt{6}}$ |
| | | | | | -1/2 | $\dfrac{\psi_{21-1}^{\alpha}}{\sqrt{3}}$ | $\dfrac{\psi_{210}^{\alpha}}{\sqrt{6}}$ | $-\dfrac{i\psi_{310}^{\alpha}}{\sqrt{6}}$ | $-\dfrac{i\psi_{31-1}^{\alpha}}{\sqrt{3}}$ |
| | | 3/2 | 1 | 3 | 3/2 | $\dfrac{\psi_{211}^{\alpha}}{\sqrt{2}}$ | 0 | 0 | $-\dfrac{i\psi_{311}^{\alpha}}{\sqrt{2}}$ |
| | | | | | 1/2 | $\dfrac{\psi_{210}^{\alpha}}{\sqrt{3}}$ | $\dfrac{\psi_{211}^{\alpha}}{\sqrt{6}}$ | $-\dfrac{i\psi_{311}^{\alpha}}{\sqrt{6}}$ | $-\dfrac{i\psi_{310}^{\alpha}}{\sqrt{3}}$ |
| | | | | | -1/2 | $-\dfrac{\psi_{21-1}^{\alpha}}{\sqrt{6}}$ | $\dfrac{\psi_{210}^{\alpha}}{\sqrt{3}}$ | $-\dfrac{i\psi_{310}^{\alpha}}{\sqrt{3}}$ | $\dfrac{i\psi_{31-1}^{\alpha}}{\sqrt{6}}$ |
| | | | | | -3/2 | 0 | $-\dfrac{\psi_{21-1}^{\alpha}}{\sqrt{2}}$ | $\dfrac{i\psi_{31-1}^{\alpha}}{\sqrt{2}}$ | 0 |
| 3 | 0 | 1/2 | 1 | 4 | 1/2 | $\dfrac{\psi_{300}^{\alpha}}{\sqrt{2}}$ | 0 | 0 | $-\dfrac{i\psi_{400}^{\alpha}}{\sqrt{2}}$ |
| | | | | | -1/2 | 0 | $\dfrac{\psi_{300}^{\alpha}}{\sqrt{2}}$ | $-\dfrac{i\psi_{400}^{\alpha}}{\sqrt{2}}$ | 0 |

| | | | | | | | | |
|---|---|---|---|---|---|---|---|---|
| | 1 | 1/2 | -1 | 4 | 1/2 | $-\dfrac{\psi^\alpha_{310}}{\sqrt{6}}$ | $\dfrac{\psi^\alpha_{311}}{\sqrt{3}}$ | $-\dfrac{i\psi^\alpha_{411}}{\sqrt{3}}$ | $\dfrac{i\psi^\alpha_{410}}{\sqrt{6}}$ |
| | | | | | -1/2 | $\dfrac{\psi^\alpha_{31-1}}{\sqrt{3}}$ | $\dfrac{\psi^\alpha_{310}}{\sqrt{6}}$ | $-\dfrac{i\psi^\alpha_{410}}{\sqrt{6}}$ | $-\dfrac{i\psi^\alpha_{41-1}}{\sqrt{3}}$ |
| | | 3/2 | 1 | 4 | 3/2 | $\dfrac{\psi^\alpha_{311}}{\sqrt{2}}$ | 0 | 0 | $-\dfrac{i\psi^\alpha_{411}}{\sqrt{2}}$ |
| | | | | | 1/2 | $\dfrac{\psi^\alpha_{310}}{\sqrt{3}}$ | $\dfrac{\psi^\alpha_{311}}{\sqrt{6}}$ | $-\dfrac{i\psi^\alpha_{411}}{\sqrt{6}}$ | $-\dfrac{i\psi^\alpha_{410}}{\sqrt{3}}$ |
| | | | | | -1/2 | $-\dfrac{\psi^\alpha_{31-1}}{\sqrt{6}}$ | $\dfrac{\psi^\alpha_{310}}{\sqrt{3}}$ | $-\dfrac{i\psi^\alpha_{410}}{\sqrt{3}}$ | $\dfrac{i\psi^\alpha_{41-1}}{\sqrt{6}}$ |
| | | | | | -3/2 | 0 | $-\dfrac{\psi^\alpha_{31-1}}{\sqrt{2}}$ | $\dfrac{i\psi^\alpha_{41-1}}{\sqrt{2}}$ | 0 |
| | 2 | 3/2 | -1 | 4 | 3/2 | $-\dfrac{\psi^\alpha_{321}}{\sqrt{10}}$ | $\sqrt{\dfrac{2}{5}}\psi^\alpha_{322}$ | $-\sqrt{\dfrac{2}{5}}i\psi^\alpha_{422}$ | $\dfrac{i\psi^\alpha_{421}}{\sqrt{10}}$ |
| | | | | | 1/2 | $-\dfrac{\psi^\alpha_{320}}{\sqrt{5}}$ | $\sqrt{\dfrac{3}{10}}\psi^\alpha_{321}$ | $-\sqrt{\dfrac{3}{10}}i\psi^\alpha_{421}$ | $\dfrac{i\psi^\alpha_{420}}{\sqrt{5}}$ |
| | | | | | -1/2 | $\sqrt{\dfrac{3}{10}}\psi^\alpha_{32-1}$ | $\dfrac{\psi^\alpha_{320}}{\sqrt{5}}$ | $-\dfrac{i\psi^\alpha_{420}}{\sqrt{5}}$ | $-\sqrt{\dfrac{3}{10}}i\psi^\alpha_{42-1}$ |
| | | | | | -3/2 | $-\sqrt{\dfrac{2}{5}}\psi^\alpha_{32-2}$ | $-\dfrac{\psi^\alpha_{32-1}}{\sqrt{10}}$ | $\dfrac{i\psi^\alpha_{42-1}}{\sqrt{10}}$ | $\sqrt{\dfrac{2}{5}}i\psi^\alpha_{42-2}$ |
| | | 5/2 | 1 | 4 | 5/2 | $\dfrac{\psi^\alpha_{322}}{\sqrt{2}}$ | 0 | 0 | $-\dfrac{i\psi^\alpha_{422}}{\sqrt{2}}$ |
| | | | | | 3/2 | $\sqrt{\dfrac{2}{5}}\psi^\alpha_{321}$ | $\dfrac{\psi^\alpha_{322}}{\sqrt{10}}$ | $-\dfrac{i\psi^\alpha_{422}}{\sqrt{10}}$ | $-\sqrt{\dfrac{2}{5}}i\psi^\alpha_{421}$ |

| | 1/2 | $\sqrt{\dfrac{3}{10}}\psi^{\alpha}_{320}$ | $\dfrac{\psi^{\alpha}_{321}}{\sqrt{5}}$ | $-\dfrac{i\psi^{\alpha}_{421}}{\sqrt{5}}$ | $-\sqrt{\dfrac{3}{10}}i\psi^{\alpha}_{420}$ |
|---|---|---|---|---|---|
| | -1/2 | $-\dfrac{\psi^{\alpha}_{32-1}}{\sqrt{5}}$ | $\sqrt{\dfrac{3}{10}}\psi^{\alpha}_{320}$ | $-\sqrt{\dfrac{3}{10}}i\psi^{\alpha}_{420}$ | $\dfrac{i\psi^{\alpha}_{42-1}}{\sqrt{5}}$ |
| | -3/2 | $\dfrac{\psi^{\alpha}_{32-2}}{\sqrt{10}}$ | $-\sqrt{\dfrac{2}{5}}\psi^{\alpha}_{32-1}$ | $\sqrt{\dfrac{2}{5}}i\psi^{\alpha}_{42-1}$ | $-\dfrac{i\psi^{\alpha}_{42-2}}{\sqrt{10}}$ |
| | -5/2 | 0 | $\dfrac{\psi^{\alpha}_{32-2}}{\sqrt{2}}$ | $-\dfrac{i\psi^{\alpha}_{42-2}}{\sqrt{2}}$ | 0 |

Table 4. The exponential type spinor orbitals in position space for $s=\frac{3}{2}$, $1 \leq n \leq 3$, $0 \leq l \leq n-1$, $|l-s| \leq j \leq l+s$, $-j \leq m \leq j$, $t=2(j-l)$ and $\tilde{n}=n+|t|$

| n | l | j | t | $\tilde{n}$ | m | $\tilde{\Psi}_{nljm}^{\alpha 3/2}$ | | | | | | | |
|---|---|---|---|---|---|---|---|---|---|---|---|---|---|
| 1 | 0 | 3/2 | 3 | 4 | 3/2 | $\frac{\psi_{100}^\alpha}{\sqrt{2}}$ | 0 | 0 | 0 | 0 | 0 | 0 | $-\frac{i\psi_{400}^\alpha}{\sqrt{2}}$ |
|   |   |   |   |   | 1/2 | 0 | $\frac{\psi_{100}^\alpha}{\sqrt{2}}$ | 0 | 0 | 0 | 0 | $-\frac{i\psi_{400}^\alpha}{\sqrt{2}}$ | 0 |
|   |   |   |   |   | −1/2 | 0 | 0 | $\frac{\psi_{100}^\alpha}{\sqrt{2}}$ | 0 | 0 | $\frac{i\psi_{400}^\alpha}{\sqrt{2}}$ | 0 | 0 |
|   |   |   |   |   | −3/2 | 0 | 0 | 0 | $\frac{\psi_{100}^\alpha}{\sqrt{2}}$ | $-\frac{i\psi_{400}^\alpha}{\sqrt{2}}$ | 0 | 0 | 0 |
| 2 | 0 | 3/2 | 3 | 5 | 3/2 | $\frac{\psi_{200}^\alpha}{\sqrt{2}}$ | 0 | 0 | 0 | 0 | 0 | 0 | $-\frac{i\psi_{500}^\alpha}{\sqrt{2}}$ |
|   |   |   |   |   | 1/2 | 0 | $\frac{\psi_{200}^\alpha}{\sqrt{2}}$ | 0 | 0 | 0 | 0 | $-\frac{i\psi_{500}^\alpha}{\sqrt{2}}$ | 0 |
|   |   |   |   |   | −1/2 | 0 | 0 | $\frac{\psi_{200}^\alpha}{\sqrt{2}}$ | 0 | 0 | $\frac{i\psi_{500}^\alpha}{\sqrt{2}}$ | 0 | 0 |
|   |   |   |   |   | −3/2 | 0 | 0 | 0 | $\frac{\psi_{200}^\alpha}{\sqrt{2}}$ | $-\frac{i\psi_{500}^\alpha}{\sqrt{2}}$ | 0 | 0 | 0 |
|   | 1 | 3/2 | 1 | 3 | 3/2 | $-\sqrt{\frac{3}{10}}\psi_{210}^\alpha$ | $\frac{\psi_{211}^\alpha}{\sqrt{5}}$ | 0 | $\sqrt{\frac{3}{10}}\psi_{210}^\alpha$ | $\frac{i\psi_{311}^\alpha}{\sqrt{5}}$ | 0 | $-\frac{i\psi_{311}^\alpha}{\sqrt{5}}$ | $\sqrt{\frac{3}{10}}i\psi_{310}^\alpha$ |
|   |   |   |   |   | 1/2 | $\frac{\psi_{21-1}^\alpha}{\sqrt{5}}$ | $-\frac{\psi_{210}^\alpha}{\sqrt{30}}$ | $\frac{2\psi_{211}^\alpha}{\sqrt{15}}$ | 0 | $-\frac{i\psi_{311}^\alpha}{\sqrt{5}}$ | $\frac{2i\psi_{311}^\alpha}{\sqrt{15}}$ | $\frac{i\psi_{310}^\alpha}{\sqrt{30}}$ | $-\frac{i\psi_{31-1}^\alpha}{\sqrt{5}}$ |
|   |   |   |   |   | −1/2 | 0 | $\frac{2\psi_{21-1}^\alpha}{\sqrt{15}}$ | $\frac{\psi_{210}^\alpha}{\sqrt{30}}$ | $\frac{\psi_{211}^\alpha}{\sqrt{5}}$ | $\frac{i\psi_{311}^\alpha}{\sqrt{5}}$ | $-\frac{i\psi_{310}^\alpha}{\sqrt{30}}$ | $-\frac{2i\psi_{31-1}^\alpha}{\sqrt{15}}$ | 0 |
|   |   |   |   |   | −3/2 | 0 | 0 | $\frac{\psi_{21-1}^\alpha}{\sqrt{5}}$ | $\sqrt{\frac{3}{10}}\psi_{210}^\alpha$ | $-\sqrt{\frac{3}{10}}i\psi_{310}^\alpha$ | $-\frac{i\psi_{31-1}^\alpha}{\sqrt{5}}$ | 0 | 0 |

| | | | | | | | | | | | |
|---|---|---|---|---|---|---|---|---|---|---|---|
| | | 5/2 | 5 | 5/2 | $\frac{\psi^\alpha_{211}}{\sqrt{2}}$ | 0 | 0 | 0 | 0 | 0 | $-\frac{i\psi^\alpha_{511}}{\sqrt{2}}$ |
| | | | | 3/2 | $\frac{\psi^\alpha_{210}}{\sqrt{5}}$ | $\sqrt{\frac{3}{10}}\psi^\alpha_{211}$ | 0 | 0 | 0 | $-\sqrt{\frac{3}{10}}i\psi^\alpha_{511}$ | $-\frac{i\psi^\alpha_{510}}{\sqrt{5}}$ |
| | | | | 1/2 | $-\frac{\psi^\alpha_{21-1}}{2\sqrt{5}}$ | $\frac{3}{\sqrt{10}}\psi^\alpha_{210}$ | $\frac{1}{2}\sqrt{\frac{3}{5}}\psi^\alpha_{211}$ | $\frac{\psi^\alpha_{211}}{2\sqrt{5}}$ | $-\frac{i\psi^\alpha_{511}}{2\sqrt{5}}$ | $-\sqrt{\frac{3}{10}}i\psi^\alpha_{510}$ | $\frac{i\psi^\alpha_{51-1}}{2\sqrt{5}}$ |
| | | | | -1/2 | 0 | $-\frac{1}{2}\sqrt{\frac{3}{5}}\psi^\alpha_{21-1}$ | $\frac{3}{\sqrt{10}}\psi^\alpha_{210}$ | $\frac{\psi^\alpha_{210}}{\sqrt{5}}$ | $-\frac{i\psi^\alpha_{510}}{\sqrt{5}}$ | $-\sqrt{\frac{3}{10}}i\psi^\alpha_{51-1}$ | 0 |
| | | | | -3/2 | 0 | 0 | $-\sqrt{\frac{3}{10}}\psi^\alpha_{21-1}$ | $-\frac{\psi^\alpha_{21-1}}{\sqrt{2}}$ | $\frac{i\psi^\alpha_{51-1}}{\sqrt{2}}$ | $\frac{1}{2}\sqrt{\frac{3}{5}}i\psi^\alpha_{51-1}$ | 0 |
| | | | | -5/2 | 0 | 0 | 0 | 0 | 0 | 0 |
| | 0 | 3 | 6 | 3/2 | $\frac{\psi^\alpha_{300}}{\sqrt{2}}$ | $\frac{\psi^\alpha_{300}}{\sqrt{2}}$ | 0 | 0 | 0 | 0 | 0 |
| | | | | 1/2 | 0 | 0 | $\frac{\psi^\alpha_{300}}{\sqrt{2}}$ | $-\frac{i\psi^\alpha_{600}}{\sqrt{2}}$ | $\frac{i\psi^\alpha_{600}}{\sqrt{2}}$ | $\frac{i\psi^\alpha_{600}}{\sqrt{2}}$ | $-\frac{i\psi^\alpha_{600}}{\sqrt{2}}$ |
| | | | | -1/2 | 0 | 0 | 0 | 0 | 0 | 0 | 0 |
| | | | | -3/2 | 0 | 0 | $\frac{\psi^\alpha_{300}}{\sqrt{2}}$ | 0 | 0 | 0 | 0 |
| 3 | 1 | 3/2 | 4 | 3/2 | $-\sqrt{\frac{3}{10}}\psi^\alpha_{310}$ | $\frac{\psi^\alpha_{311}}{\sqrt{5}}$ | $\frac{2\psi^\alpha_{311}}{\sqrt{15}}$ | 0 | 0 | $-\frac{i\psi^\alpha_{411}}{\sqrt{5}}$ | $\sqrt{\frac{3}{10}}i\psi^\alpha_{410}$ |
| | | | | 1/2 | $\frac{\psi^\alpha_{31-1}}{\sqrt{5}}$ | $-\frac{\psi^\alpha_{310}}{\sqrt{30}}$ | $\frac{2i\psi^\alpha_{411}}{\sqrt{15}}$ | 0 | 0 | $\frac{i\psi^\alpha_{410}}{\sqrt{30}}$ | $-\frac{i\psi^\alpha_{41-1}}{\sqrt{5}}$ |
| | | | -1/2 | 0 | $\frac{2\psi^\alpha_{31-1}}{\sqrt{15}}$ | $\frac{\psi^\alpha_{310}}{\sqrt{30}}$ | $\frac{\psi^\alpha_{311}}{\sqrt{5}}$ | $-\frac{i\psi^\alpha_{411}}{\sqrt{5}}$ | $-\frac{i\psi^\alpha_{410}}{\sqrt{30}}$ | $-\frac{2i\psi^\alpha_{41-1}}{\sqrt{15}}$ | 0 |
| | | | -3/2 | 0 | 0 | $\frac{\psi^\alpha_{31-1}}{\sqrt{5}}$ | $\frac{3}{\sqrt{10}}\psi^\alpha_{310}$ | $-\frac{3}{\sqrt{10}}i\psi^\alpha_{410}$ | $-\frac{i\psi^\alpha_{41-1}}{\sqrt{5}}$ | 0 | 0 |
| 5/2 | 3 | 6 | 5/2 | $\frac{\psi^\alpha_{311}}{\sqrt{2}}$ | 0 | 0 | 0 | 0 | 0 | $-\frac{i\psi^\alpha_{611}}{\sqrt{2}}$ |
| | | | 3/2 | $\frac{\psi^\alpha_{310}}{\sqrt{5}}$ | 0 | $\sqrt{\frac{3}{10}}\psi^\alpha_{311}$ | 0 | $-\sqrt{\frac{3}{10}}i\psi^\alpha_{611}$ | 0 | $-\frac{3}{\sqrt{10}}i\psi^\alpha_{611}$ | $-\frac{i\psi^\alpha_{610}}{\sqrt{5}}$ |
| | | | 1/2 | $-\frac{\psi^\alpha_{31-1}}{2\sqrt{5}}$ | $\sqrt{\frac{3}{10}}\psi^\alpha_{310}$ | $\frac{1}{2}\sqrt{\frac{3}{5}}\psi^\alpha_{311}$ | 0 | $-\frac{1}{2}\sqrt{\frac{3}{5}}i\psi^\alpha_{611}$ | $-\frac{3}{\sqrt{10}}i\psi^\alpha_{610}$ | $-\frac{3}{\sqrt{10}}i\psi^\alpha_{611}$ | $\frac{i\psi^\alpha_{61-1}}{2\sqrt{5}}$ |
| | | | -1/2 | 0 | $-\frac{1}{2}\sqrt{\frac{3}{5}}\psi^\alpha_{31-1}$ | 0 | $\frac{\psi^\alpha_{311}}{2\sqrt{5}}$ | $-\frac{3}{\sqrt{10}}i\psi^\alpha_{610}$ | $-\frac{3}{\sqrt{10}}i\psi^\alpha_{611}$ | $\frac{1}{2}\sqrt{\frac{3}{5}}i\psi^\alpha_{61-1}$ | 0 |
| | | | -3/2 | 0 | 0 | $-\frac{3}{\sqrt{10}}\psi^\alpha_{31-1}$ | $\frac{\psi^\alpha_{310}}{\sqrt{5}}$ | $\frac{i\psi^\alpha_{610}}{\sqrt{5}}$ | $\frac{3}{\sqrt{10}}i\psi^\alpha_{61-1}$ | 0 | 0 |
| | | | -5/2 | 0 | 0 | 0 | $-\frac{i\psi^\alpha_{31-1}}{\sqrt{2}}$ | $\frac{i\psi^\alpha_{61-1}}{\sqrt{2}}$ | 0 | 0 | 0 |
| 2 | 3/2 | -1 | 4 | 3/2 | $\frac{\psi^\alpha_{320}}{\sqrt{10}}$ | $-\frac{\psi^\alpha_{321}}{\sqrt{5}}$ | $\frac{\psi^\alpha_{322}}{\sqrt{5}}$ | $-\frac{i\psi^\alpha_{422}}{\sqrt{5}}$ | $\frac{i\psi^\alpha_{421}}{\sqrt{5}}$ | $-\frac{i\psi^\alpha_{420}}{\sqrt{10}}$ |
| | | | 1/2 | $-\frac{\psi^\alpha_{32-1}}{\sqrt{5}}$ | $-\frac{\psi^\alpha_{320}}{\sqrt{10}}$ | 0 | $\frac{\psi^\alpha_{321}}{\sqrt{5}}$ | $-\frac{i\psi^\alpha_{421}}{\sqrt{5}}$ | 0 | $\frac{i\psi^\alpha_{420}}{\sqrt{10}}$ | $-\frac{i\psi^\alpha_{42-1}}{\sqrt{5}}$ |
| | | | -1/2 | $\frac{\psi^\alpha_{32-2}}{\sqrt{5}}$ | 0 | $-\frac{\psi^\alpha_{320}}{\sqrt{10}}$ | $\frac{\psi^\alpha_{320}}{\sqrt{10}}$ | 0 | $\frac{i\psi^\alpha_{420}}{\sqrt{10}}$ | 0 | $-\frac{i\psi^\alpha_{42-2}}{\sqrt{5}}$ |
| | | | -3/2 | 0 | $\frac{\psi^\alpha_{32-2}}{\sqrt{5}}$ | 0 | $\frac{\psi^\alpha_{32-1}}{\sqrt{5}}$ | $-\frac{i\psi^\alpha_{42-1}}{\sqrt{5}}$ | $\frac{i\psi^\alpha_{42-2}}{\sqrt{5}}$ | 0 |
| 5/2 | 1 | 4 | 5/2 | $-\sqrt{\frac{3}{14}}\psi^\alpha_{321}$ | $\sqrt{\frac{2}{7}}\psi^\alpha_{322}$ | 0 | 0 | 0 | 0 | $-\sqrt{\frac{2}{7}}i\psi^\alpha_{422}$ | $\sqrt{\frac{3}{14}}i\psi^\alpha_{421}$ |
| | | | 3/2 | $-\frac{3\psi^\alpha_{320}}{\sqrt{35}}$ | $\frac{\psi^\alpha_{321}}{\sqrt{70}}$ | $2\sqrt{\frac{2}{35}}\psi^\alpha_{322}$ | 0 | 0 | 0 | $\frac{i\psi^\alpha_{421}}{\sqrt{70}}$ | $\frac{3i\psi^\alpha_{420}}{\sqrt{35}}$ |
| | | | 1/2 | $\frac{3}{2}\sqrt{\frac{3}{35}}\psi^\alpha_{32-1}$ | $-\frac{3}{\sqrt{70}}\psi^\alpha_{320}$ | $\frac{1}{2}\sqrt{\frac{5}{7}}\psi^\alpha_{321}$ | $\sqrt{\frac{3}{35}}\psi^\alpha_{322}$ | 0 | 0 | $-\frac{3}{2}\sqrt{\frac{3}{35}}i\psi^\alpha_{422}$ | $-\frac{3}{2}\sqrt{\frac{3}{35}}i\psi^\alpha_{42-1}$ |
| | | | -1/2 | $-\sqrt{\frac{3}{35}}\psi^\alpha_{32-2}$ | $\frac{1}{2}\sqrt{\frac{5}{7}}\psi^\alpha_{32-1}$ | $\frac{3}{\sqrt{70}}\psi^\alpha_{320}$ | $\frac{3}{2}\sqrt{\frac{3}{35}}\psi^\alpha_{321}$ | 0 | 0 | $-\frac{3}{2}\sqrt{\frac{3}{35}}i\psi^\alpha_{421}$ | $\sqrt{\frac{3}{35}}i\psi^\alpha_{42-2}$ |
| | | | -3/2 | 0 | $-2\sqrt{\frac{2}{35}}\psi^\alpha_{32-}$ | $\frac{\psi^\alpha_{32-1}}{\sqrt{70}}$ | $\frac{3\psi^\alpha_{320}}{\sqrt{35}}$ | $-\frac{3}{\sqrt{35}}i\psi^\alpha_{422}$ | $\frac{3i\psi^\alpha_{420}}{\sqrt{35}}$ | $\frac{1}{2}\sqrt{\frac{5}{7}}i\psi^\alpha_{42-1}$ | $\sqrt{\frac{3}{35}}i\psi^\alpha_{42-2}$ |
| | | | -5/2 | 0 | 0 | $-\sqrt{\frac{2}{7}}\psi^\alpha_{32-2}$ | $-\sqrt{\frac{3}{14}}\psi^\alpha_{32-1}$ | $\sqrt{\frac{3}{14}}i\psi^\alpha_{42-1}$ | $\sqrt{\frac{3}{14}}i\psi^\alpha_{42-1}$ | $2\sqrt{\frac{2}{35}}i\psi^\alpha_{42-2}$ | 0 |
| 7/2 | 3 | 6 | 7/2 | $\frac{\psi^\alpha_{322}}{\sqrt{2}}$ | $\sqrt{\frac{3}{14}}\psi^\alpha_{322}$ | 0 | 0 | 0 | 0 | 0 | $-\frac{i\psi^\alpha_{622}}{\sqrt{2}}$ |
| | | | 5/2 | $\sqrt{\frac{2}{7}}\psi^\alpha_{321}$ | $\sqrt{\frac{2}{7}}\psi^\alpha_{321}$ | $\frac{\psi^\alpha_{322}}{\sqrt{14}}$ | 0 | 0 | $-\frac{i\psi^\alpha_{622}}{\sqrt{14}}$ | $-\sqrt{\frac{3}{14}}i\psi^\alpha_{622}$ | $-\sqrt{\frac{2}{7}}i\psi^\alpha_{621}$ |
| | | | 3/2 | $\frac{\psi^\alpha_{320}}{\sqrt{7}}$ | $\frac{3\psi^\alpha_{320}}{\sqrt{35}}$ | $\sqrt{\frac{6}{35}}\psi^\alpha_{321}$ | $-\frac{i\psi^\alpha_{622}}{\sqrt{70}}$ | $-\sqrt{\frac{6}{35}}i\psi^\alpha_{621}$ | $-\sqrt{\frac{3}{14}}i\psi^\alpha_{621}$ | $-\sqrt{\frac{2}{7}}i\psi^\alpha_{621}$ | $-\frac{i\psi^\alpha_{620}}{\sqrt{7}}$ |
| | | | 1/2 | $-\sqrt{\frac{2}{35}}\psi^\alpha_{32-1}$ | $\frac{3\psi^\alpha_{320}}{\sqrt{35}}$ | $\sqrt{\frac{6}{35}}\psi^\alpha_{321}$ | 0 | $-\sqrt{\frac{6}{35}}i\psi^\alpha_{621}$ | $-\sqrt{\frac{6}{35}}i\psi^\alpha_{621}$ | $-\frac{3i\psi^\alpha_{620}}{\sqrt{35}}$ | $\sqrt{\frac{2}{35}}i\psi^\alpha_{62-1}$ |

| | | | | | | | |
|---|---|---|---|---|---|---|---|
| -1/2 | $\frac{\psi^\alpha_{32-2}}{\sqrt{70}}$ | $-\sqrt{\frac{6}{35}}\psi^\alpha_{32-1}$ | $\frac{3\psi^\alpha_{320}}{\sqrt{35}}$ | $\sqrt{\frac{2}{35}}\psi^\alpha_{321}$ | $-\sqrt{\frac{2}{35}}i\psi^\alpha_{621}$ | $-\frac{3i\psi^\alpha_{620}}{\sqrt{35}}$ | $\sqrt{\frac{6}{35}}i\psi^\alpha_{62-1}$ | $-\frac{i\psi^\alpha_{62-2}}{\sqrt{70}}$ |
| -3/2 | 0 | $\frac{\psi^\alpha_{32-2}}{\sqrt{14}}$ | $-\sqrt{\frac{2}{7}}\psi^\alpha_{32-1}$ | $\frac{\psi^\alpha_{320}}{\sqrt{7}}$ | $-\frac{i\psi^\alpha_{620}}{\sqrt{7}}$ | $\sqrt{\frac{2}{7}}i\psi^\alpha_{62-1}$ | $-\frac{i\psi^\alpha_{62-2}}{\sqrt{14}}$ | 0 |
| -5/2 | 0 | 0 | $\sqrt{\frac{3}{14}}\psi^\alpha_{32-2}$ | $-\sqrt{\frac{2}{7}}\psi^\alpha_{32-1}$ | $\sqrt{\frac{2}{7}}i\psi^\alpha_{62-1}$ | $-\sqrt{\frac{3}{14}}i\psi^\alpha_{62-2}$ | 0 | 0 |
| -7/2 | 0 | 0 | 0 | $-\frac{\psi^\alpha_{32-2}}{\sqrt{2}}$ | $-\frac{i\psi^\alpha_{62-2}}{\sqrt{2}}$ | 0 | 0 | 0 |